\pdfoutput=1
\documentclass[12pt,titlepage]{article}

\font\grande=cmr9.5 scaled \magstep4
\font\medio=cmr9.5 scaled \magstep2
\outer\def\beginsection#1\par{\medbreak\bigskip
      \message{#1}\leftline{\bf#1}\nobreak\medskip
\vskip-\parskip
      \noindent}
\usepackage{graphicx} 
\begin{document}
\bibliographystyle {unsrt}

\titlepage

\begin{flushright}
\end{flushright}

\vspace{1cm}
\begin{center}
{\grande Squeezed relic photons beyond the horizon}\\
\vspace{1cm}
 Massimo Giovannini 
 \footnote{Electronic address: massimo.giovannini@cern.ch} \\
\vspace{1cm}
{{\sl Department of Physics, 
Theory Division, CERN, 1211 Geneva 23, Switzerland }}\\
\vspace{0.5cm}
{{\sl INFN, Section of Milan-Bicocca, 20126 Milan, Italy}}
\vspace*{1cm}
\end{center}
 \vskip 0.3cm
\centerline{\medio  Abstract}
\vskip 0.1cm
Owing to the analogy with the ordinary photons in the visible range of the 
electromagnetic spectrum, the Glauber theory is generalized to address the
quantum coherence of the gauge field fluctuations 
parametrically amplified during an inflationary stage 
of expansion. The first and second  degrees of quantum coherence 
of relic photons are then computed beyond the effective horizon 
defined by the evolution of the susceptibility. In the zero-delay limit the Hanbury Brown-Twiss 
correlations exhibit a super-Poissonian statistics which 
is however different from the conventional results of the single-mode approximation 
customarily employed, in quantum optics, to classify the coherence 
properties of visible light. While in the case of large-scale curvature perturbations the 
degrees of quantum coherence coincide with the naive expectation of the 
single-mode approximation, the net degree of second-order 
coherence computed for the relic photons diminishes thanks to the effect of the polarizations. 
We suggest that the Hanbury Brown-twiss correlations are probably the only tool to 
assess the quantum or classical origin of the large-scale 
magnetic fluctuations and of the corresponding curvature perturbations. 

\noindent

\vspace{5mm}
\vfill
\newpage
\renewcommand{\theequation}{1.\arabic{equation}}
\setcounter{equation}{0}
\section{Introduction}
\label{sec1}
The squeezed states of optical photons arise  
in a number of diverse physical situations all related
(directly or indirectly) to the quantum theory of the parametric 
amplification \cite{mandel}. The formulation of the quantum theory 
of optical coherence  \cite{glauber1,glauber1a,glauber2,titulaer} 
paved the way for the first quantum description of parametric 
amplification \cite{mollow}. Since then various complementary descriptions of quantum 
amplifiers have been developed through the years \cite{stoler,yuen,hollenhorst,caves} 
both in the context of single-mode and two-mode squeezed states 
(see also \cite{revsq1,revsq2} for an incomplete list of review articles on the subject).

After the seminal discoveries of the COBE satellite \cite{cobe} (later 
confirmed and extended by the WMAP experiment \cite{wmap1,wmap2}) 
it became gradually clear that the early Universe itself could be seen, 
from the physical viewpoint, as an effective quantum amplifier. 
Consequently the applications of quantum optical techniques to the analysis 
of large-scale inhomogeneities has been firstly suggested by 
Grishchuk and collaborators in a class of problems involving the 
evolution of the tensor and scalar modes 
of the four-dimensional geometry \cite{gr1,gr2,gr3}. 
Neither the tensor \cite{gr0} nor the scalar \cite{luk,KS,chibisov} inhomogeneities 
of a conformally flat geometry  are invariant under Weyl rescaling 
of the four-dimensional metric. The lack of Weyl invariance implies then the 
formation of squeezed states of the relic gravitons and of the relic phonons 
\cite{gr1,gr2,gr3} (see also \cite{gr4} for a  review article). 
The key physical assumption behind these attempts rests on 
the quantum mechanical nature of the initial conditions of the 
large-scale inhomogeneities, as suggested long ago by Sakharov \cite{sak} 
even prior to the formulation of the conventional 
inflationary paradigms. 

The quantum theory of parametric amplification has been later applied to the case 
of relic photons \cite{mg1} where the quantum optical analogy is even more compelling: 
in this case it is precisely the time variation of the susceptibility that plays the 
role of the laser pump often employed for the direct experimental  
preparation of the squeezed states in various classes of nonlinear materials 
(see e.g. \cite{mandel,revsq1,revsq2} and also \cite{LK}).  The quantum theory of parametric amplification 
of the relic photons (but also of the relic gravitons and relic phonons) is useful for treating 
the problem of initial data but it becomes essential for analyzing the higher-order correlations of the 
large-scale fluctuations, as the quantum optical analogy clearly suggests. 

There are some who argue that we have already an accurate control of the protoinflationary dynamics; 
along this prespective a consistent model suffices for claiming that the large-scale fluctuations have a 
quantum origin. In spite of this belief, it would be nice (and probably even mandatory) to develop a more 
objective set of sufficient criteria enabling us to infer the quantum origin of large-scale fluctuations of any spin 
from some sort of observational evidence. The first idea coming to mind, in this 
respect, it is to analyze the quantum coherence of the fluctuations in the spirit 
of the Glauber theory  \cite{glauber1,glauber1a,glauber2}.
Only by looking at the higher-order correlations we shall be able, at least in principle,  
 to establish if the large-scale curvature perturbations have a classical or a quantum 
origin  as speculated by Sakharov \cite{sak}.

A first step along this direction relies on the idea of studying (and eventually measuring)
 the correlation functions of the intensities of the curvature perturbations  rather than the correlations of the 
 corresponding amplitudes \cite{mg2}.  This concept has been originally proposed by Hanbury Brown and  Twiss 
 \cite{HBT0} and their analysis of the intensity correlations is often dubbed Hanbury Brown-Twiss (HBT) interferometry as 
 opposed to the standard Young-type interference where only amplitudes (rather than intensities) are concerned.
The applications of the HBT  effect range from stellar astronomy \cite{HBT0} (see also \cite{HBT01})  to subatomic physics \cite{revs} 
where the interference of the intensities has been used to determine the hadron fireball dimensions \cite{cocconi} 
corresponding, in rough terms, to the linear size of the interaction region in proton-proton collisions.

In this paper the quantum theory of optical coherence is applied to the scrutiny of 
the statistical properties of the relic photons produced thanks to the pumping action of the susceptibility 
during an inflationary stage of expansion.  The idea is to define the Glauber
correlation functions and to focus the attention on their large-scale limit. 
The first and second degrees of quantum coherence correspond, in the quantum optical 
analogy, to the Young interferometry and to the HBT interferometry. 
In the zero-time delay limit the degree of second-order coherence (conventionally denoted by  $g^{(2)}$ in quantum optics \cite{mandel})
can be used to infer the statistical properties of the quantum state.
In the standard lore, based on the so-called single mode approximation \cite{mandel}, $g^{(2)} \to 1 $ 
for a coherent state (also referred to as the Poissonian limit because 
of the well known statistical properties of the coherent states). Conversely in the chaotic (or thermal) case we 
would have $g^{(2)}\to 2$; finally in the case of two-mode squeezed states $g^{(2)}\to 3$ signalling a 
super-Poissonian but also superchaotic statistics. By comparing the the Hanbury-Brown Twiss correlations 
computed in the scalar case (and, more precisely, for the large-scale curvature 
fluctuations) with the case of relic photons we find specific physical differences which are traced back to the role of the polarizations. 

The plan of the present paper is the following. In section \ref{sec2} we shall discuss the squeezed states of the relic photons.
In section \ref{sec3} the essentials of the Glauber approach will be introduced. The large-scale limits 
of the correlation functions will be studied in section \ref{sec4}. In section \ref{sec5} the physical meaning of the 
degrees of quantum coherence will be specifically computed and contrasted with the single-mode approximation.
Section \ref{sec6} contains our concluding remarks. To avoid digressions, various useful details have been relegated to the appendix.

\renewcommand{\theequation}{2.\arabic{equation}}
\setcounter{equation}{0}
\section{Squeezed states of relic photons}
\label{sec2}
The conformally invariant coupling of the Abelian gauge fields is broken in different situations 
that can be usefully recapitulated in terms of the general action \cite{action}:
\begin{equation}
S = \int d^4 x \, \sqrt{- g} \biggl[ {\mathcal M}_{\sigma}^{\rho}(\varphi,\psi)
Y_{\rho\alpha}\, Y^{\sigma\alpha} - {\mathcal N}_{\sigma}^{\rho}(\varphi,\psi) \widetilde{Y}_{\rho\alpha}\, \widetilde{Y}^{\sigma\alpha}
\biggr],
\label{action}
\end{equation}
where  ${\mathcal M}_{\sigma}^{\rho}(\varphi,\psi)$ and ${\mathcal N}_{\sigma}^{\rho}(\varphi,\psi)$ may depend on a 
number of different scalar fields and on their covariant derivatives. In a complementary perspective they can be 
constructed directly from fluid variables (i.e. fluid velocities, vorticities and shear).
In spite of their specific form, when ${\mathcal M}_{\sigma}^{\rho}\neq {\mathcal N}_{\sigma}^{\rho}$ the system is characterized by different 
electric and magnetic susceptibilities; in this situation Eq. (\ref{action}) includes, as a special case, the derivative couplings arising in the relativistic theory of Casimir-Polder and Van der Waals interactions \cite{such}. We shall be assuming, consistently with the observations, that the evolution of the large-scale magnetic fields takes place in a conformally flat background geometry $\overline{g}_{\mu\nu} = a^2(\tau)\eta_{\mu\nu}$ where 
$\eta_{\mu\nu}$ denotes the Minkowski metric, $a(\tau)$ is the scale factor and $\tau$ denotes the conformal time coordinate.
If ${\mathcal M}_{\sigma}^{\rho}\neq {\mathcal N}_{\sigma}^{\rho}$ the comoving electric and magnetic fields obey the following set of equations:  
\begin{eqnarray}
&& \vec{\nabla} \times \biggl( \sqrt{\Lambda_{B}} \vec{B} \biggr) = \partial_{\tau} \biggl( \sqrt{\Lambda_{E}} \vec{E}\biggr),
\label{onebb}\\
&& \vec{\nabla} \times \biggl(\frac{\vec{E}}{\sqrt{\Lambda_{E}}}\biggr) + \partial_{\tau} \biggl(\frac{\vec{B}}{\sqrt{\Lambda_{B}}}\biggr) =0,
\label{twob}\\
&& \vec{\nabla} \cdot \biggl(\frac{\vec{B}}{\sqrt{\Lambda_{B}}}\biggr)=0,\qquad \vec{\nabla}\cdot ( \sqrt{\Lambda_{E}}\, \vec{E} )=0.
\label{threeb}
\end{eqnarray}
The electric and magnetic couplings are, respectively, $g_{E} = (4\pi/\Lambda_{E})^{1/2}$ and $g_{B} = (4\pi/\Lambda_{B})^{1/2}$.
Under the exchange and inversion of the susceptibilities ($\sqrt{\Lambda}_{E} \to 1/\sqrt{\Lambda_{B}}$ and $\sqrt{\Lambda_{B}} \to 1/\sqrt{\Lambda_{E}}$)  or of the corresponding couplings (i.e.  $g_{E} \to 1/g_{B}$ and $g_{B} \to 1/g_{E}$) Eqs. (\ref{onebb}), (\ref{twob}) and (\ref{threeb}) 
maintain the same form provided  the electric and magnetic fields are also exchanged as $\vec{E} \to - \vec{B}$ and  $\vec{B} \to \vec{E}$. 
Even if the discussion can be carried on in the general case, we shall be focussing our attention on the simplest situation, namely the one where 
${\mathcal M}_{\sigma}^{\rho} = {\mathcal N}_{\sigma}^{\rho} = (\lambda/2) \delta_{\sigma}^{\rho}$. In this instance
Eqs. (\ref{onebb}), (\ref{twob}) and (\ref{threeb})  become\footnote{This situation corresponds to various models of magnetogenesis \cite{rev1,rev2} discussed in the past \cite{DT1,DT2,DT3}. See also \cite{DT4,DT4a} for a recent observation leading to an interesting class of magnetogenesis models not described by Eq. (\ref{action}). } 
\begin{equation}
 \vec{E}^{\prime} +  {\mathcal F} \vec{E} = \vec{\nabla}\times \vec{B}, \qquad \vec{B}^{\prime} - {\mathcal F} \vec{B} =  - \vec{\nabla}\times \vec{E},
\label{second}
\end{equation}
where ${\mathcal F} = \chi^{\,\prime}/\chi$, $\chi =  \sqrt{\lambda}$ is the susceptibility and the prime denotes a derivation with respect to the conformal time coordinate.  The components of the Abelian field strength of Eq. (\ref{action}) are defined as $Y^{0i} = e^{i}/a^2$ and $Y^{ij}= -\epsilon^{ijk} b_{k}/a^2$.  
The canonical electric 
and magnetic fields appearing in Eq. (\ref{second}) are then given by $\vec{B} = a^2 \, \sqrt{\lambda}\, \vec{b}$ and  $\vec{E} = a^2 \, \sqrt{\lambda}\, \vec{e}$. Note that the two equations appearing in Eq. (\ref{second}) are left invariant by the duality transformations \cite{duality1} $\chi\to 1/\chi$ (i.e. ${\mathcal F} \to - {\mathcal F}$) provided $\vec{E} \to - \vec{B}$ and  $\vec{B} \to \vec{E}$. The continuous evolution of ${\mathcal F}$ will define an effective horizon for the gauge modes related to $\vec{E}$ and $\vec{B}$.

 In time-dependent (conformally flat) backgrounds the Coulomb gauge (i.e. $Y_{0} =0$  and $\vec{\nabla}\cdot \vec{Y} =0$)  is preserved (unlike the Lorentz gauge condition) under a conformal rescaling of the metric. For the quantum mechanical description of the problem we can therefore start with the canonical Hamiltonian (see appendix \ref{APPA} for a derivation)
\begin{equation}
\hat{H}(\tau) = \sum_{\alpha} \int d^{3}k \biggl[ \frac{k}{2} (\hat{a}_{\vec{k}\, \alpha}^{\dagger} \, \hat{a}_{\vec{k}\, \alpha} + \hat{a}_{-\vec{k}\, \alpha} \, \hat{a}_{-\vec{k}\, \alpha}^{\dagger})
+ \xi\, \hat{a}_{\vec{k}\, \alpha}^{\dagger} \hat{a}_{-\vec{k}\, \alpha}^{\dagger}  + \xi^{\ast} \,  \hat{a}_{-\vec{k}\, \alpha}\, \hat{a}_{\vec{k}\, \alpha}\biggr].
\label{one}
\end{equation} 
where $\xi = i {\mathcal F}/2$. 
Equation (\ref{one}) is reminiscent of the toy model of parametric amplifier analyzed, for the first time by Mollow and Glauber \cite{mollow}. The free part of Eq. (\ref{one}) and the two components of the interacting Hamiltonian satisfy the usual commutation relations of the $SU(1,1)$ Lie algebra, as 
we shall see in a moment.  Equation (\ref{one}) describes an interacting Bose gas at zero temperature. In this case the free 
Hamiltonian corresponds to the kinetic energy while the interaction terms account for the two-body collisions with small momentum transfer  \cite{fetter,solomon}. 

The Hamiltonian (\ref{one}) is invariant under duality that transforms $\chi$ in its inverse, i.e.
$\chi\to 1/\chi$. Under this transformation we have that ${\mathcal F} \to - {\mathcal F}$ while the creation and annihilation 
operators transform as:
\begin{eqnarray}
&& \hat{a}_{\vec{k}\, \alpha} \to i k \, \hat{a}_{-\vec{k}\,\alpha}^{\dagger}, \qquad  \hat{a}_{-\vec{k}\, \alpha} \to \frac{i}{k} \, \hat{a}_{\vec{k}\,\alpha}^{\dagger}
\label{onea}\\
&&  \hat{a}_{\vec{k}\, \alpha}^{\dagger} \to - \frac{i}{k}\, \hat{a}_{-\vec{k}\, \alpha}, \qquad  \hat{a}_{-\vec{k}\, \alpha}^{\dagger} \to - i k \, \hat{a}_{\vec{k}\, \alpha}.
\label{oneb}
\end{eqnarray}
Recalling the notations discussed in appendix \ref{APPA},  the Fourier representation of the field operators and of the momenta 
\begin{equation}
\hat{{\mathcal A}}_{\vec{k}\, \alpha} = \frac{1}{\sqrt{2 k}} ( \hat{a}_{\vec{k}\, \alpha} + \hat{a}_{-\vec{k}\, \alpha}^{\dagger}),\qquad \hat{\pi}_{\vec{k}\, \alpha} = - i \sqrt{\frac{k}{2}} ( \hat{a}_{\vec{k}\, \alpha} - \hat{a}_{-\vec{k}\, \alpha}^{\dagger}),
\label{onec}
\end{equation}
transform as 
\begin{equation}
\hat{{\mathcal A}}_{\vec{k}\, \alpha} \to \frac{\hat{\pi}_{\vec{k}\, \alpha}}{k}, \qquad \hat{\pi}_{\vec{k}\, \alpha} \to - k \hat{{\mathcal A}}_{\vec{k}\, \alpha}
\label{oned}
\end{equation}
if we use Eqs. (\ref{onea}) and (\ref{oneb}). In the present discussion the vacuum corresponds to the state minimizing the Hamiltonian at the onset of the dynamical evolution. This state can be  explicitly constructed by diagonalizing the Hamiltonian in terms of an appropriate canonical 
transformation. A similar procedure is used to derive the ground state wavefunction of an interacting Bose gas at zero 
temperature \cite{fetter,solomon}.

The evolution of $\hat{a}_{\vec{k}\, \alpha}$ and $\hat{a}^{\dagger}_{\vec{k}\, \alpha}$ can be obtained 
from Eq. (\ref{one}) and from the evolution equations in the Heisenberg description: 
\begin{eqnarray}
\frac{d \hat{a}_{\vec{p}\, \alpha}}{d\tau} &=&  i [ \hat{H},\,\hat{a}_{\vec{p}\, \alpha}] = - i\, p \, \hat{a}_{\vec{p}\, \alpha} - 2 i \, \xi \hat{a}_{- \vec{p}\, \alpha}^{\dagger},
\nonumber\\
 \frac{d \hat{a}_{\vec{p}\, \alpha}^{\dagger}}{d\tau} &=& i [ \hat{H},\,\hat{a}^{\dagger}_{\vec{p}\, \alpha}] = i\, p \, \hat{a}^{\dagger}_{\vec{p}\, \alpha} + 2 i \, \xi^{\ast} \hat{a}_{-\vec{p}\, \alpha}.
\label{two}
\end{eqnarray}
The formal solution of Eq. (\ref{two}) is
\begin{eqnarray}
\hat{a}_{\vec{p},\, \alpha}(\tau,\,\tau_{i}) &=& u_{p}(\tau) \,\,\hat{b}_{\vec{p}\, \alpha}(\tau_{i}) - v_{p}(\tau) \,\, \hat{b}_{- \vec{p}\, \alpha}^{\dagger}(\tau_{i}),
\nonumber\\
\hat{a}_{-\vec{p},\, \alpha}^{\dagger}(\tau,\,\tau_{i}) &=& u_{p}^{\ast}(\tau)\,\, \hat{b}_{-\vec{p}\, \alpha}^{\dagger}(\tau_{i}) - v_{p}^{\ast}(\tau) \,\,\hat{b}_{\vec{p}\, \alpha}(\tau_{i}),
\label{three}
\end{eqnarray}
where $u_{p}(\tau)$ and $v_{p}(\tau)$  satisfy $|u_{p}(\tau)|^2 - |v_{p}(\tau)|^2 =1$.  From Eq. (\ref{two}) the equations obeyed by $u_{p}$ and $v_{p}$ can be written as:
\begin{equation}
\frac{d u_{p}}{d\tau} = - i p \,u_{p} - {\mathcal F} v_{p}^{\ast}, \qquad \frac{d v_{p}}{d\tau} = - i p\, v_{p} - {\mathcal F} u_{p}^{\ast}.
\label{five}
\end{equation}
The solution for the evolution equations of $u_{p}(\tau)$ and $v_{p}(\tau)$ can be obtained in two 
complementary regions, namely for the wavelengths larger than the effective horizon (i.e. $p/{\mathcal F} \ll 1$) and for 
wavelengths shorter than the effective horizon (i.e. $p/{\mathcal F} \gg 1$). In the short wavelength 
region the solutions of Eq. (\ref{five}) are plane waves $e^{\pm i p \tau}$ while in the long 
wavelength regime the solution becomes:
\begin{eqnarray}
u_{k}(\tau) &=& A_{k}(\tau,\tau_{ex}) \, u_{k}(\tau_{ex}) + B_{k}^{*}(\tau,\tau_{ex}) \, v_{k}^{*}(\tau_{ex}),
\label{U1}\\
v_{k}^{*}(\tau) &=& B_{k}(\tau,\tau_{ex})  \, u_{k}(\tau_{ex}) + A_{k}^{*}(\tau,\tau_{ex})\,v_{k}^{*}(\tau_{ex}),
\label{V1}
\end{eqnarray}
where $A_{k}(\tau,\tau_{ex})$ and $B_{k}(\tau,\tau_{ex})$ are given by:
\begin{eqnarray}
A_{k}(\tau,\tau_{ex}) &=& \frac{\chi(\tau)}{2 \chi_{ex}} \biggl[ 1 + i\, {\mathcal I}_{B}(\tau_{ex},\tau)\biggr] +  \frac{\chi_{ex}}{2 \chi(\tau)} \biggl[ 1 - i\, {\mathcal I}_{E}(\tau_{ex},\tau)\biggr],
\label{AA1}\\
B_{k}(\tau,\tau_{ex}) &=& \frac{\chi_{ex}}{2 \chi(\tau)} \biggl[ 1 - i\, {\mathcal I}_{E}(\tau_{ex},\tau)\biggr]  - \frac{\chi(\tau)}{2 \chi_{ex}} \biggl[ 1 + i\, {\mathcal I}_{B}(\tau_{ex},\tau)\biggr].
\label{BB1}
\end{eqnarray}
The two dimensionless integrals ${\mathcal I}_{B}(\tau_{ex},\tau)$ and ${\mathcal I}_{E}(\tau_{ex},\tau)$ are given by 
\begin{equation}
 {\mathcal I}_{B}(\tau_{ex},\tau) = k \int_{\tau_{ex}}^{\tau} \frac{\chi_{ex}^2}{\chi(\tau^{\prime})} d\tau^{\prime}, \qquad  
 {\mathcal I}_{E}(\tau_{ex},\tau) = k \int_{\tau_{ex}}^{\tau} \frac{\chi(\tau^{\prime})}{\chi_{ex}^2} d\tau^{\prime}.
 \label{INT1}
 \end{equation}
Thanks to Eqs. (\ref{U1}) and (\ref{V1}) the initial conditions for the evolution can be directly expressed at $\tau_{ex}$ and can be written 
in terms of the values of the mode functions at the corresponding epoch 
(i.e. $ u_{p}(\tau_{ex}) \equiv \overline{u}_{p}$ and $v^{*}_{p}(\tau_{ex}) \equiv \overline{v}^{*}_{p}$).

We can now remark that  the two complex functions $u_{p}(\tau)$ and $v_{p}(\tau)$ (subjected to the constraint  $|u_{p}(\tau)|^2 - |v_{p}(\tau)|^2 =1$) can the be parametrized in terms of three real functions. The evolution of $u_{k}$ and $v_{k}$ can then be rephrased in terms of the so-called squeezing parameters \cite{mandel,revsq1,revsq2} (see also \cite{stoler,yuen,hollenhorst,caves}):
\begin{equation}
u_{p}=  e^{- i \varphi_{p}}\cosh{r_{p}}, \qquad v_{p}= e^{- i (\varphi_{p} -\gamma_{p})} \sinh{r_{p}},
\label{four}
\end{equation}
where  $\varphi_{p}$, $r_{p}$ and $\gamma_{p}$ are all functions of the 
conformal time coordinate $\tau$ even if the arguments of the functions will be dropped for the sake of conciseness.
Using Eq. (\ref{four}),  Eq. (\ref{three}) can be rewritten as:
\begin{eqnarray} 
\hat{a}_{\vec{p}\, \alpha} &=& e^{- i \varphi_{p}}\biggl[ \cosh{r_{p}}\, \hat{b}_{\vec{p}\,\alpha} - e^{i \gamma_{p}} \sinh{r_{p}} \hat{b}^{\dagger}_{-\vec{p}\,\alpha}\biggr],
\nonumber\\
\hat{a}_{-\vec{p}\, \alpha}^{\dagger} &=&  e^{i \varphi_{p}}\biggl[ \cosh{r_{p}}\, \hat{b}_{-\vec{p}\, \alpha}^{\dagger} - e^{-i \gamma_{p}} \sinh{r_{p}}\hat{b}_{\vec{p}\,\alpha}\biggr].
\label{AA20}
\end{eqnarray}
Equation (\ref{AA20}) can be swiftly obtained by considering a single $\vec{p}$-mode and by noticing that 
the operators ${\mathcal K}_{\pm}$ and ${\mathcal K}_{0}$ obey the commutation relations of the $SU(1,1)$ Lie algebra:
\begin{eqnarray}
{\mathcal K}_{+} = \hat{b}_{1}^{\dagger} \,\hat{b}_{2}^{\dagger},\qquad {\mathcal K}_{-} = \hat{b}_{1}\, \hat{b}_{2},\qquad 
{\mathcal K}_{0} = \frac{1}{2}\biggl[ \hat{b}_{1}^{\dagger}\, \hat{b}_{1} + \hat{b}_{2} \,\hat{b}_{2}^{\dagger}\biggr].
\label{AA21}
\end{eqnarray}
Using the the standard Campbell-Baker-Hausdorff theorem \cite{mandel,schum}, Eq. (\ref{AA21}) implies 
\begin{equation}
\hat{a} = \Sigma^{\dagger}(\zeta) \, \Xi^{\dagger}(\varphi) b_{1} \Xi(\varphi) \Sigma(\zeta) = 
e^{- i \varphi}\biggl[ \cosh{r}\, \hat{b}_{1} - e^{i \gamma} \sinh{r} \hat{b}^{\dagger}_{2}\,\biggr],
\label{AA22}
\end{equation}
where $\Xi(\varphi)$ and $\Sigma(\zeta)$ (with $\zeta= r e^{i\gamma}$) are, respectively, the rotation operator and the two-mode 
squeezing operators defined in terms of the generators of the $SU(1,1)$ Lie algebra:
\begin{equation}
\Xi(\varphi) = \exp{[ - i \varphi (\hat{b}_{1}^{\dagger} \hat{b}_{1} + \hat{b}_{2} \hat{b}_{2}^{\dagger})]}, \qquad \Sigma(\zeta)= 
\exp{[ \zeta^{*}\, \hat{b}_{1} \,\hat{b}_{2} - \zeta \, \hat{b}_{2}^{\dagger}\, \hat{b}_{1}^{\dagger}]}.
\label{AA23}
\end{equation}
These two operators describe the evolution of the states in the Schr\"odinger representation; their use has been 
pioneered by Grishchuk and Sidorov \cite{gr1} (see also \cite{mg1} in the case of the relic photons). 
Using Eq. (\ref{AA20}) into  Eqs. (\ref{five}), the evolution of the squeezing amplitude $r_{k}$ and of the phase $\varphi_{p}$ becomes:
\begin{equation}
r_{p}^{\prime} = - {\mathcal F} \cos{\alpha_{p}}, \qquad 
\varphi_{p}^{\prime} = p + {\mathcal F} \sin{\alpha_{p}}\, \tanh{r_{p}},
\label{AA24}
\end{equation}
where $\alpha_{p} = 2 \varphi_{p} - \gamma_{p}$ and the relation between $\gamma_{p}^{\prime}$ and $\varphi_{p}^{\prime}$ is given by:
\begin{equation}
\gamma_{p}^{\prime} = \varphi_{p}^{\prime} - p - {\mathcal F} \frac{\sin{\alpha_{p}}}{\tanh{r_{p}}}.
\label{AA24a}
\end{equation}
By combining Eqs. (\ref{AA24}) and (\ref{AA24a}) it is immediate to obtain
\begin{equation}
\alpha_{p}^{\prime} = 2 p + 2 {\mathcal F} \frac{\sin{\alpha_{p}}}{\tanh{2 r_{p}}}.
\label{AA24b}
\end{equation} 

\renewcommand{\theequation}{3.\arabic{equation}}
\setcounter{equation}{0}
\section{Glauber description of quantum coherence}
\label{sec3}
\subsection{General form of the Glauber correlation function}

The statistical properties of any quantum state and its degrees of quantum coherence 
can be used to reconstruct, at least in partially, the physical nature of the source 
\cite{mandel,glauber1,glauber1a,glauber2}. 
In quantum optics the Glauber theory is often used in an exclusive manner: the statistical 
properties of visible light are reduced to the study of a single mode of the field. This is what goes 
under the name of single-mode approximation. Conversely, in 
the analysis of the large-scale cosmological fluctuations of  different spin, 
a more inclusive approach is needed since the correlation functions contain all the modes of the field. 
In its most general form the Glauber correlation function can be written as \cite{glauber1,glauber2}:
\begin{eqnarray}
&& {\mathcal G}_{i_{1}, \,.\,.\,.\,i_{n}, \, i_{n+1},\, .\,.\,.\,, i_{n +m} } ^{(n,m)}(x_{1}, \,.\,.\,.\,x_{n}, \, x_{n+1},\, .\,.\,.\,, x_{n +m}) 
\nonumber\\
&& = \mathrm{Tr}\biggl[ \hat{\rho} \, \hat{{\mathcal A}}_{i_{1}}^{(-)}(x_{1})\,.\,.\,.\, \hat{{\mathcal A}}_{i_{n}}^{(-)}(x_{n})
\, \hat{{\mathcal A}}_{i_{n+1}}^{(+)}(x_{n+1})\,.\,.\,.\,\hat{{\mathcal A}}_{i_{n+m}}^{(+)}(x_{n+m})\biggr],
\label{corr1}
\end{eqnarray}
where $x_{i} \equiv (\vec{x}_{i}, \, \tau_{i})$ and $\hat{\rho}$ is the density operator representing the (generally mixed) state of the field $\hat{{\mathcal A}}_{i}$. 
 The field $\hat{{\mathcal A}}_{i}(\vec{x}, \tau)$ can always be expressed as $\hat{{\mathcal A}}_{i}(x) = \hat{{\mathcal A}}_{i}^{(+)}(x) + \hat{{\mathcal A}}_{i}^{(-)}(x)$, with $\hat{{\mathcal A}}_{i}^{(+)}(x)= \hat{{\mathcal A}}_{i}^{(-)\,\dagger}(x)$.
By definition we will have that $\hat{{\mathcal A}}_{i}^{(+)}(x) |\mathrm{vac} \rangle=0$ and also that  $\langle \mathrm{vac} |\, \hat{{\mathcal A}}_{i}^{(-)}(x) =0$;  the state $|\mathrm{vac}\rangle $ denotes the vacuum. The vacuum corresponds to the state minimizing the Hamiltonian at the onset of the dynamical evolution. This state can be  explicitly constructed by diagonalizing the Hamiltonian in terms of an appropriate canonical 
transformation. A similar procedure is used to derive the ground state wavefunction of an interacting Bose gas at zero 
temperature \cite{fetter,solomon}. Provided the total duration of inflation exceeds the minimal number of about $65$ efolds, 
the vacuum initial data are the most plausible, at least in the conventional lore (see, however, Ref. \cite{mg2} for different choices 
in a related context). The correlation function defined in Eq. (\ref{corr1}) depends on the polarizations as the free indices clearly show. It is 
useful, for future convenience, to introduce the Glauber correlation function for a scalar degree of freedom.
In this case Eq. (\ref{corr1}) simply becomes:
\begin{eqnarray}
&& {\mathcal S}^{(n,m)}(x_{1}, \,.\,.\,.\,x_{n}, \, x_{n+1},\, .\,.\,.\,, x_{n +m}) 
\nonumber\\
&& = \mathrm{Tr}\biggl[ \hat{\rho} \, \hat{q}^{(-)}(x_{1})\,.\,.\,.\, \hat{q}^{(-)}(x_{n})
\, \hat{q}^{(+)}(x_{n+1})\,.\,.\,.\, \hat{q}^{(+)}(x_{n+m})\biggr].
\label{corr1a}
\end{eqnarray}
The quantum field $\hat{q}(x)$ defines, for instance, the normalized curvature 
perturbations on comoving orthogonal hypersurfaces.

It is relevant to remark that  Eq. (\ref{corr1}) (and, similarly, Eq. (\ref{corr1a}))  contain an operator that can be written as:
\begin{equation}
\hat{O}_{i_{1}, \,.\,.\,.\,i_{n}} (x_{1}, \,.\,.\,.\,x_{n}) = \hat{{\mathcal A}}_{i_{1}}^{(-)}(x_{1})\,.\,.\,.\, \hat{{\mathcal A}}_{i_{n}}^{(-)}(x_{n})
\, \hat{{\mathcal A}}_{i_{1}}^{(+)}(x_{1})\,.\,.\,.\,\hat{{\mathcal A}}_{i_{n}}^{(+)}(x_{n}).
\label{corr3}
\end{equation}
The operator of Eq. (\ref{corr3}) is needed to describe $n$-fold delayed coincidence measurements of the field at the space-time points $(x_{1}, \,.\,.\,.\,x_{n})$.
If $|\, b\rangle$ is the state before the measurement and $|\, a\rangle$ is the state after the measurement, the matrix element corresponding 
to the absorption of the quanta of $\hat{{\mathcal A}}_{i}$  at each detector and at given times is $\langle a\, | \hat{{\mathcal A}}_{i_{1}}^{(+)}(x_{1})\,.\,.\,.\,\hat{{\mathcal A}}_{i_{n}}^{(+)}(x_{n})|\, b\rangle$. The rate at which such absorptions occur, summed over the final states, is therefore proportional to \cite{mandel,glauber1,glauber2}: 
\begin{eqnarray}
&& \sum_{a} \biggl|\langle a\, | \hat{{\mathcal A}}_{i_{1}}^{(+)}(x_{1})\,.\,.\,.\,\hat{{\mathcal A}}_{i_{n}}^{(+)}(x_{n})|\, b\rangle \biggr|^2 = 
\nonumber\\
&& \sum_{a} \langle b|\hat{{\mathcal A}}_{i_{1}}^{(-)}(x_{1})\,.\,.\,.\,\hat{{\mathcal A}}_{i_{n}}^{(-)}(x_{n})| a\rangle \langle a| 
 \hat{{\mathcal A}}_{i_{1}}^{(+)}(x_{1})\,.\,.\,.\,\hat{{\mathcal A}}_{i_{n}}^{(+)}(x_{n}) |b \rangle = \langle b| \hat{O} |b \rangle,
\label{corr4}
\end{eqnarray}
where the second equality of Eq. (\ref{corr4}) follows from the completeness relation. 

\subsection{Symmetric form of the correlation function}

According to Eq. (\ref{corr4}), when $\langle b| \hat{O} |b \rangle$ is averaged over the ensemble of the initial states of the system it becomes identical with Eq. (\ref{corr1}) for $x_{n + r} = x_{r}$ (with $r= 1, 2, \,.\,.\,., n$ and $n=m$). Since this is the case that will be studied hereunder, we shall denote the symmetric form of the Glauber correlation function as:
\begin{eqnarray}
&& {\mathcal G}_{i_{1}, \,.\,.\,.\,i_{n},\,i_{n+1}, \,.\,.\,.\,i_{2 n} }^{(n)}(x_{1}, \,.\,.\,.\,x_{n}, \, x_{n+1},\, .\,.\,.\,, x_{2n})
\nonumber\\
&&  = \mathrm{Tr}\biggl[ \hat{\rho} \, \hat{{\mathcal A}}_{i_{1}}^{(-)}(x_{1})\,.\,.\,.\, \hat{{\mathcal A}}_{i_{n}}^{(-)}(x_{n})
\, \hat{{\mathcal A}}_{i_{n+1}}^{(+)}(x_{n+1})\,.\,.\,.\,\hat{{\mathcal A}}_{i_{2n}}^{(+)}(x_{2n})\biggr].
\label{corr5}
\end{eqnarray}
Thanks to Eq. (\ref{corr5}), the coherence properties of the quantum field $\hat{{\mathcal A}}_{i}(x)$ can be discussed by introducing the normalized version of  the $n$-point Glauber function \cite{glauber1,glauber2}:
\begin{equation}
g^{(n)}_{i_{1}, \,.\,.\,.\,i_{n},\,i_{n+1}, \,.\,.\,.\,i_{2 n} }(x_{1}, \,.\,.\,.\,x_{n}, \, x_{n+1},\, .\,.\,.\,, x_{2n}) = \frac{{\mathcal G}_{i_{1}, \,.\,.\,.\,i_{n}} ^{(n)}(x_{1}, \,.\,.\,.\,x_{n}, \, x_{n+1},\, .\,.\,.\,, x_{2n})}{\sqrt{\Pi_{j =1}^{2 n}\, {\mathcal G}_{i_{j} i_{j}}^{(1)}(x_{j}, x_{j})}}.
\label{corr6}
\end{equation}
While, by definition, $|g^{(1)}_{i_{1} \,i_{2}}(x_{1},\, x_{2})| \leq 1$ the higher order correlators are not restricted in absolute value as it happens for $g^{(1)}(x_{1}, x_{2})$. A fully  coherent field  must therefore satisfy the following necessary condition \cite{mandel,glauber1,glauber2}:
\begin{equation}
g^{(n)}_{i_{1}, \,.\,.\,.\,i_{n},\,i_{n+1}, \,.\,.\,.\,i_{2 n} } (x_{1}, \,.\,.\,.\,x_{n}, \, x_{n+1},\, .\,.\,.\,, x_{2n}) = 1,
\label{corr6a}
\end{equation}
for $ n = 1,\, 2,\, 3,\, .\,.\,.$.  If only a limited number of normalized correlation functions will satisfy Eq. (\ref{corr6a}) we shall speak about partial 
coherence.  The degrees of first- and second-order coherence can be written as:
\begin{eqnarray}
&& g_{i_{1} \, i_{2}}^{(1)}(x_{1}, x_{2}) = \frac{{\mathcal G}^{(1)}_{i_{1}\, i_{2}}(x_{1}, x_{2})}{\sqrt{{\mathcal G}^{(1)}_{i_{1} i_{1}} (x_{1}, x_{1})\, {\mathcal G}^{(1)}_{i_{2} i_{2}} (x_{2}, x_{2})}},
\label{g1}\\
&& g^{(2)}_{i_{1}\, i_{2}\, i_{3}\, i_{4}}(x_{1}, x_{2}, x_{3}, x_{4}) = \frac{{\mathcal G}_{i_{1}\, i_{2}\, i_{3}\, i_{4}}^{(2)}(x_{1}, x_{2}, x_{3}, x_{4})}{\sqrt{{\mathcal G}_{i_{1}\, i_{1}}^{(1)}(x_{1}, x_{1})\, {\mathcal G}_{i_{2}\, i_{2}}^{(1)}(x_{2}, x_{2})\, {\mathcal G}_{i_{3}\, i_{3}}^{(1)}(x_{3}, x_{3})\,{\mathcal G}_{i_{4}\, i_{4}}^{(1)}(x_{4}, x_{4})}},
\label{g2}
\end{eqnarray}
where, in agreement with the general definitions of Eq. (\ref{corr1}), the correlation functions appearing in Eqs. (\ref{g1}) and (\ref{g2}) 
are given by:
\begin{eqnarray}
{\mathcal G}^{(1)}_{i_{1}\, i_{2}}(x_{1}, x_{2}) &=& \langle \hat{{\mathcal A}}^{(-)}_{i_{1}}(x_{1}) \hat{{\mathcal A}}^{(+)}_{i_{2}}(x_{2}) \rangle,
\nonumber\\
{\mathcal G}_{i_{1}\, i_{2}\, i_{3}\, i_{4}}^{(2)}(x_{1}, x_{2}, x_{3}, x_{4})&=& \langle \hat{{\mathcal A}}^{(-)}_{i_{1}}(x_{1}) \hat{{\mathcal A}}^{(-)}_{i_{2}}(x_{2}) \hat{{\mathcal A}}^{(+)}_{i_{3}}(x_{3}) \hat{{\mathcal A}}^{(+)}_{i_{4}}(x_{4})\rangle.
\label{ex1}
\end{eqnarray}
In a similar manner it is possible to define, for instance the third- and fourth-order degrees of coherence
\begin{eqnarray}
g^{(3)}(x_{1},\,x_{2},\, x_{3},\, x_{4},\,x_{5},\, x_{6}) &=&  \frac{{\mathcal G}^{(3)}(x_{1},\,x_{2},\, x_{3},\, x_{4},\,x_{5},\,x_{6})}{\sqrt{\prod_{i =1}^{6} {\mathcal G}^{(1)}(x_{i},\, x_{i})}},
\label{g3}\\
g^{(4)}(x_{1},\, x_{2},\, x_{3},\, x_{4},\,x_{5}, x_{6},\, x_{7},\,x_{8}) &=&  \frac{{\mathcal G}^{(4)}(x_{1},\,x_{2},\, x_{3},\, x_{4},\,x_{5},\,x_{6},\,x_{7},\,x_{8})}{\sqrt{\prod_{i \,=\,1}^{8} {\mathcal G}^{(1)}(x_{i},\, x_{i})}},
\label{g4}
\end{eqnarray}
where, for the sake of simplicity, we just suppressed the polarization indices.
If $g^{(1)}(x_{1},\, x_{2})= 1$ and $g^{(2)}(x_{1},\, x_{2},\, x_{3},\, x_{4}) =1$ 
(but $g^{(3)}(x_{1},\, x_{2},\, x_{3},\, x_{4},\, x_{5},\,x_{6}) \neq 1$)  the quantum field is second-order coherent. 
We shall be interested in the first and second degrees of coherence 
even if It has been recently suggested, in quantum optical applications, that  the degree of second-order coherence might not always be  sufficient to specify completely the statistical properties of the radiation field \cite{opt1,opt2,opt3,opt4}. 

\subsection{Electric and magnetic correlation functions}

The Glauber correlation function of Eq. (\ref{corr5}) has been originally defined not in terms 
of the vector potentials but rather using the electric fields:
\begin{eqnarray}
&& {\mathcal E}^{(n)}{i_{1}, \,.\,.\,.\,i_{n},\,i_{n+1}, \,.\,.\,.\,i_{2 n} } (x_{1}, \,.\,.\,.\,x_{n}, \, x_{n+1},\, .\,.\,.\,, x_{2n})
\nonumber\\
&&  = \mathrm{Tr}\biggl[ \hat{\rho} \, \hat{E}_{i_{1}}^{(-)}(x_{1})\,.\,.\,.\, \hat{E}_{i_{n}}^{(-)}(x_{n})
\, \hat{E}_{i_{n+1}}^{(+)}(x_{1})\,.\,.\,.\,\hat{E}_{i_{n}}^{(+)}(x_{2 n})\biggr].
\label{corrE}
\end{eqnarray}
From Eq. (\ref{corrE}) the corresponding degrees of second-order coherence can also be defined. Equation (\ref{corr5}) has been instead 
proposed as basic correlator in the approach of Mandel and Wolf \cite{mandel}. Both approaches are somewhat convenient 
for applications to questions relating to photoelectric detection of light fluctuations. 
In the present context exactly the same discussion can be carried on in the case of the magnetic correlator defined as:
\begin{eqnarray}
&& {\mathcal B}^{(n)}_{i_{1}, \,.\,.\,.\,i_{n},\,i_{n+1}, \,.\,.\,.\,i_{2 n} } (x_{1}, \,.\,.\,.\,x_{n}, \, x_{n+1},\, .\,.\,.\,, x_{2n})
\nonumber\\
&&  = \mathrm{Tr}\biggl[ \hat{\rho} \, \hat{B}_{i_{1}}^{(-)}(x_{1})\,.\,.\,.\, \hat{B}_{i_{n}}^{(-)}(x_{n})
\, \hat{B}_{i_{1}}^{(+)}(x_{n+1})\,.\,.\,.\,\hat{B}_{i_{n}}^{(+)}(x_{2 n})\biggr].
\label{corrB}
\end{eqnarray}
From Eqs. (\ref{corrE}) and (\ref{corrB})  the normalized degrees of quantum coherence can be easily defined 
from the expressions already derived\footnote{The electric and magnetic correlators give coincident results for the 
degrees of quantum coherence as we shall explicitly show in the next section.
This property should be contrasted with what happens for the magnetic and electric power spectra (see also appendix \ref{APPC}).
The reason for this occurrence is that the degrees of quantum coherence, by construction, are sensitive to the properties of the quantum state.} using Eq. (\ref{corr5}). 

The degree of first-order coherence of Eq. (\ref{g1}) appears naturally in the Young two-slit experiment and whenever the degree of first-order coherence is equal to $1$ the visibility is maximized \cite{mandel}.
The degree of second-order coherence of Eq. (\ref{g2}) enters the discussion of the Hanbury Brown-Twiss effect \cite{HBT0} and its different applications 
ranging from stellar interferometry \cite{mandel} to high-energy physics \cite{revs,cocconi}. The degree of second-order coherence 
arises naturally when discussing the correlations of the intensities of the fields $\hat{A}_{i}$, $\hat{E}_{i}$ and $\hat{B}_{i}$.  Notice that 
the intensity correlators relevant to the HBT interferometry can be easily obtained from Eqs. (\ref{corr6}) and (\ref{g2}) by identifying the space-time points as 
follows:
\begin{equation}
x_{1} \equiv x_{n+1},\qquad x_{2} \equiv x_{n+2},\qquad .\,.\,.\, \qquad x_{n} \equiv x_{2 n}.
\label{ident}
\end{equation}
In this case the original Glauber correlator will effectively be a function of $n$ points and and it will describe the correlation 
of $n$ intensities. The same observation can be made in the case of Eqs. (\ref{corrE}) and (\ref{corrB}). The explicit 
expressions of the HBT correlators can then be written from Eqs. (\ref{corr1}), (\ref{corrE}) and (\ref{corrB}) 
with the help of Eq. (\ref{ident}):
\begin{eqnarray} 
{\mathcal G}^{(2)}(x_{1}, x_{2}) &=&  \sum_{i_{1}\, i_{2}} \langle \hat{{\mathcal A}}^{(-)}_{i_{1}}(x_{1}) \hat{{\mathcal A}}^{(-)}_{i_{2}}(x_{2}) \hat{{\mathcal A}}^{(+)}_{i_{1}}(x_{1}) \hat{{\mathcal A}}^{(+)}_{i_{2}}(x_{2})\rangle,
\label{HBTG}\\
{\mathcal E}^{(2)}(x_{1}, x_{2}) &=&  \sum_{i_{1}\, i_{2}} \langle \hat{E}^{(-)}_{i_{1}}(x_{1}) \hat{E}^{(-)}_{i_{2}}(x_{2}) \hat{E}^{(+)}_{i_{1}}(x_{1}) \hat{E}^{(+)}_{i_{2}}(x_{2})\rangle,
\label{HBTE}\\
{\mathcal B}^{(2)}(x_{1}, x_{2}) &=&   \sum_{i_{1}\, i_{2}} \langle \hat{B}^{(-)}_{i_{1}}(x_{1}) \hat{B}^{(-)}_{i_{2}}(x_{2}) \hat{B}^{(+)}_{i_{1}}(x_{1}) \hat{B}^{(+)}_{i_{2}}(x_{2})\rangle,
\label{HBTB}
\end{eqnarray}
where the sum over repeated indices is pleonastic since the usual convention of sum over repeated indices has been adopted throughout.
Nonetheless the explicit form of Eqs. (\ref{HBTG}), (\ref{HBTG}) and (\ref{HBTB}) can be revealing when compared with the explicit 
form of Eq. (\ref{corr1a}) in the case of HBT correlations:
\begin{equation}
{\mathcal S}^{(2)}(x_{1}, x_{2}) = \langle  \hat{q}^{(-)}(x_{1}) \hat{q}^{(-)}(x_{2}) \hat{q}^{(+)}(x_{1}) \hat{q}^{(+)}(x_{2})\rangle.
\label{HBTS}
\end{equation}
The difference between Eqs. (\ref{HBTG})--(\ref{HBTB}) and Eq. (\ref{HBTS}) will have a direct 
repercussion on the large-scale limits of the degree of quantum coherence, as we shall see in the following section.

\renewcommand{\theequation}{4.\arabic{equation}}
\setcounter{equation}{0}
\section{Quantum correlators beyond the effective horizon}
\label{sec4}
The correlation functions introduced in section \ref{sec3} will now be
 computed in the case of the squeezed quantum states associated with the Hamiltonian of Eq. (\ref{one}). 
 To avoid digressions some of the relevant details have been relegated in the appendices \ref{APPB} and \ref{APPC}.

\subsection{Explicit form of the correlators}

In the case $n=1$, Eqs. (\ref{corr5}) and (\ref{corrE})--(\ref{corrB}) give the the explicit expressions of the first-order 
correlators:
\begin{eqnarray}
{\mathcal G}^{(1)}_{ij}(x_{1}, x_{2}) &=& \frac{1}{2} \int \frac{d^{3} p}{(2\pi)^3 \, p} P_{ij}(\hat{p})\, v_{p}^{*}(\tau_{1}) v_{p}(\tau_{2}) \, e^{- i \vec{p}\cdot\vec{r}}, 
\label{firstA}\\
{\mathcal B}^{(1)}_{ij}(x_{1}, x_{2}) ={\mathcal E}^{(1)}_{ij}(x_{1}, x_{2}) &=& \frac{1}{2} \int \frac{d^{3} p}{(2\pi)^3}\, p\, P_{ij}(\hat{p})\, v_{p}^{*}(\tau_{1}) v_{p}(\tau_{2}) \, e^{- i \vec{p}\cdot\vec{r}},
\label{firstEB}
\end{eqnarray}
where $\vec{r} = \vec{x}_{1} - \vec{x}_{2}$ and $P_{ij}(\hat{p}) = \delta_{ij} - \hat{p}_{i} \hat{p}_{j}$ (with $\hat{p}_{i} = p_{i}/p$).  
The final form of the expectation values appearing in Eqs. (\ref{firstA}) and (\ref{firstEB}) can be obtained from
Eqs. (\ref{four}), (\ref{AA20}) and (\ref{B2}).
In Eqs. (\ref{firstA})--(\ref{firstEB}) we did not sum over the polarizations and
even if the previous equations hold also for $i \neq j$, the degrees of first-order coherence are actually defined 
from the traces of Eqs. (\ref{firstA})--(\ref{firstEB}):
\begin{equation}
{\mathcal G}^{(1)}(x_{1}, x_{2}) = {\mathcal G}^{(1)}_{ii}(x_{1}, x_{2}), \qquad {\mathcal B}^{(1)}(x_{1}, x_{2}) = {\mathcal B}^{(1)}_{ii}(x_{1}, x_{2}),\qquad 
{\mathcal E}^{(1)}(x_{1}, x_{2}) = {\mathcal E}^{(1)}_{ii}(x_{1}, x_{2}).
\label{trace}
\end{equation}
Within the notations Eq. (\ref{trace}), the corresponding degrees of first-order electric and magnetic coherence are, respectively,
\begin{eqnarray}
g_{{\mathcal E}}^{(1)}(x_{1},\, x_{2}) &=& \frac{{\mathcal E}^{(1)}(x_{1}, \, x_{2})}{\sqrt{{\mathcal E}^{(1)}(x_{1}, \, x_{1})} \, \sqrt{{\mathcal E}^{(1)}(x_{2}, \, x_{2})}},
\nonumber\\
g_{{\mathcal B}}^{(1)}(x_{1},\, x_{2}) &=& \frac{{\mathcal B}^{(1)}(x_{1}, \, x_{2})}{\sqrt{{\mathcal B}^{(1)}(x_{1}, \, x_{1})} \, \sqrt{{\mathcal B}^{(1)}(x_{2}, \, x_{2})}},
\label{corEB}\\
g_{{\mathcal G}}^{(1)}(x_{1},\, x_{2}) &=& \frac{{\mathcal G}^{(1)}(x_{1}, \, x_{2})}{\sqrt{{\mathcal G}^{(1)}(x_{1}, \, x_{1})} \, \sqrt{{\mathcal G}^{(1)}(x_{2}, \, x_{2})}}.
\label{corG}
\end{eqnarray}
As a consequence of Eq. (\ref{firstEB}) we also have that  $g_{\mathcal E}^{(1)}(x_{1}, x_{2}) = g_{\mathcal B}^{(1)}(x_{1}, x_{2})$. 

Equations (\ref{corr5}), (\ref{corrE}) and (\ref{corrB}) give the degree of second-order coherence when 
written in the case $n=2$. More specifically, when $n=2$  Eq. (\ref{corrB}) is given by Eq. (\ref{magnetic4}) of the appendix; 
then, after making explicit the expectation values (see Eq. (\ref{magnetic5})) the final result is: 
\begin{eqnarray}
{\mathcal B}^{(2)}(x_{1}, x_{2}, x_{3}, x_{4}) &=& \frac{1}{4} \int \frac{d^{3} p_{1}}{(2\pi)^3} p_{1} \,  \int \frac{d^{3} p_{2}}{(2\pi)^3} p_{2} 
\nonumber\\
&\times& \biggl\{ v_{p_{1}}^{*}(\tau_{1}) v_{p_{2}}^{*}(\tau_{2}) v_{p_{1}}(\tau_{3}) v_{p_{2}}(\tau_{4}) e^{- i \vec{p}_{1} \cdot(\vec{x}_{1} - \vec{x}_{3})} 
e^{- i \vec{p}_{2} \cdot(\vec{x}_{2} - \vec{x}_{4})} \, P_{ii}(\hat{p}_{1}) \, P_{ii}(\hat{p}_{2})
\nonumber\\
&+& P_{ij}(\hat{p_{1}}) \, P_{ij}(\hat{p}_{2}) \, \biggl[v_{p_{1}}^{*}(\tau_{1}) v_{p_{2}}^{*}(\tau_{2}) v_{p_{2}}(\tau_{3}) v_{p_{1}}(\tau_{4}) e^{- i \vec{p}_{1} \cdot(\vec{x}_{1} - \vec{x}_{4})} 
e^{- i \vec{p}_{2} \cdot(\vec{x}_{2} - \vec{x}_{3})} 
\nonumber\\
&+& v_{p_{1}}^{*}(\tau_{1}) u_{p_{1}}^{*}(\tau_{2}) u_{p_{2}}(\tau_{3}) v_{p_{2}}(\tau_{4}) e^{- i \vec{p}_{1} \cdot(\vec{x}_{1} - \vec{x}_{2})} 
e^{- i \vec{p}_{2} \cdot(\vec{x}_{3} - \vec{x}_{4})} \biggr]\biggr\}.
\label{B2result}
\end{eqnarray}
Had we started from Eq. (\ref{corrE}),  the same steps would have led, through Eq. (\ref{electric4}), exactly to the 
same final expression of Eq. (\ref{B2result}): in other 
words the direct calculation shows that ${\mathcal B}^{(2)}(x_{1}, x_{2}, x_{3}, x_{4}) \equiv {\mathcal E}^{(2)}(x_{1}, x_{2}, x_{3}, x_{4})$. 
For the present ends and as a preparation for the discussion of the last part of section \ref{sec5}, it is relevant to contrast Eq. (\ref{B2result}) with the 
degree of second-order coherence obtainable in the case of a scalar field \cite{mg2}. The Hamiltonian coincides, in this case, with Eq. (\ref{one}) but the sum over the 
polarizations and the polarization dependence of the creation and annihilation operators are absent. When $m = n = 2$ Eq. (\ref{corr1a}) implies:
\begin{eqnarray}
{\mathcal S}^{(2)}(x_{1}, x_{2}, x_{3}, x_{4}) &=& \frac{1}{4} \int \frac{d^{3} p_{1}}{(2\pi)^3\, p_{1}}  \,  \int \frac{d^{3} p_{2}}{(2\pi)^3\, p_{2}}
\nonumber\\
&\times& \biggl\{ v_{p_{1}}^{*}(\tau_{1}) v_{p_{2}}^{*}(\tau_{2}) v_{p_{1}}(\tau_{3}) v_{p_{2}}(\tau_{4}) e^{- i \vec{p}_{1} \cdot(\vec{x}_{1} - \vec{x}_{3})} 
e^{- i \vec{p}_{2} \cdot(\vec{x}_{2} - \vec{x}_{4})} \, \nonumber\\
&+& \, \biggl[v_{p_{1}}^{*}(\tau_{1}) v_{p_{2}}^{*}(\tau_{2}) v_{p_{1}}(\tau_{3}) v_{p_{2}}(\tau_{4}) e^{- i \vec{p}_{1} \cdot(\vec{x}_{1} - \vec{x}_{4})} 
e^{- i \vec{p}_{2} \cdot(\vec{x}_{2} - \vec{x}_{3})} 
\nonumber\\
&+& v_{p_{1}}^{*}(\tau_{1}) u_{p_{1}}^{*}(\tau_{2}) u_{p_{2}}(\tau_{3}) v_{p_{2}}(\tau_{4}) e^{- i \vec{p}_{1} \cdot(\vec{x}_{1} - \vec{x}_{2})} 
e^{- i \vec{p}_{2} \cdot(\vec{x}_{3} - \vec{x}_{4})} \biggr]\biggr\},
\label{S2result}
\end{eqnarray}
where the results of Eqs. (\ref{scalar1}) and (\ref{scalar2}) have been taken into account.
Equations (\ref{B2result}) and (\ref{S2result}) are similar but the polarizations introduce a quantitive difference which is 
even more apparent  when Eq. (\ref{B2result}) is written in explicit terms:
\begin{eqnarray}
{\mathcal B}^{(2)}(x_{1}, x_{2}, x_{3}, x_{4}) &=& {\mathcal E}^{(2)}(x_{1}, x_{2}, x_{3}, x_{4})= \int \frac{d^{3} p_{1}}{(2\pi)^3} p_{1} \,  \int \frac{d^{3} p_{2}}{(2\pi)^3} p_{2} 
\nonumber\\
&\times& \biggl\{ v_{p_{1}}^{*}(\tau_{1}) v_{p_{2}}^{*}(\tau_{2}) v_{p_{1}}(\tau_{3}) v_{p_{2}}(\tau_{4}) e^{- i \vec{p}_{1} \cdot(\vec{x}_{1} - \vec{x}_{3})} 
e^{- i \vec{p}_{2} \cdot(\vec{x}_{2} - \vec{x}_{4})} 
\nonumber\\
&+& \frac{1}{4} \biggl[ 1 + \frac{(\vec{p}_{1}\cdot\vec{p}_{2})^2}{p_{1}^2 p_{2}^2}\biggr]\biggl[v_{p_{1}}^{*}(\tau_{1}) v_{p_{2}}^{*}(\tau_{2}) v_{p_{2}}(\tau_{3}) v_{p_{1}}(\tau_{4}) e^{- i \vec{p}_{1} \cdot(\vec{x}_{1} - \vec{x}_{4})} 
e^{- i \vec{p}_{2} \cdot(\vec{x}_{2} - \vec{x}_{3})} 
\nonumber\\
&+& v_{p_{1}}^{*}(\tau_{1}) u_{p_{1}}^{*}(\tau_{2}) u_{p_{2}}(\tau_{3}) v_{p_{2}}(\tau_{4}) e^{- i \vec{p}_{1} \cdot(\vec{x}_{1} - \vec{x}_{2})} 
e^{- i \vec{p}_{2} \cdot(\vec{x}_{3} - \vec{x}_{4})} \biggr]\biggr\}.
\label{expl2A}\\
{\mathcal G}^{(2)}(x_{1}, x_{2}, x_{3}, x_{4}) &=& \int \frac{d^{3} p_{1}}{p_{1} (2\pi)^3} \,  \int \frac{d^{3} p_{2}}{p_{2}(2\pi)^3}  
\nonumber\\
&\times& \biggl\{ v_{p_{1}}^{*}(\tau_{1}) v_{p_{2}}^{*}(\tau_{2}) v_{p_{1}}(\tau_{3}) v_{p_{2}}(\tau_{4}) e^{- i \vec{p}_{1} \cdot(\vec{x}_{1} - \vec{x}_{3})} 
e^{- i \vec{p}_{2} \cdot(\vec{x}_{2} - \vec{x}_{4})} 
\nonumber\\
&+& \frac{1}{4} \biggl[ 1 + \frac{(\vec{p}_{1}\cdot\vec{p}_{2})^2}{p_{1}^2 p_{2}^2}\biggr]\biggl[v_{p_{1}}^{*}(\tau_{1}) v_{p_{2}}^{*}(\tau_{2}) v_{p_{2}}(\tau_{3}) v_{p_{1}}(\tau_{4}) e^{- i \vec{p}_{1} \cdot(\vec{x}_{1} - \vec{x}_{4})} 
e^{- i \vec{p}_{2} \cdot(\vec{x}_{2} - \vec{x}_{3})} 
\nonumber\\
&+& v_{p_{1}}^{*}(\tau_{1}) u_{p_{1}}^{*}(\tau_{2}) u_{p_{2}}(\tau_{3}) v_{p_{2}}(\tau_{4}) e^{- i \vec{p}_{1} \cdot(\vec{x}_{1} - \vec{x}_{2})} 
e^{- i \vec{p}_{2} \cdot(\vec{x}_{3} - \vec{x}_{4})} \biggr]\biggr\}.
\label{G2result}
\end{eqnarray}
While the electric and the magnetic correlators of Eqs. (\ref{corrE}) and (\ref{corrB}) lead to the same results (i.e. Eq. (\ref{expl2A})), if we use 
the vector potential as pivotal variable (as suggested, for instance, in \cite{mandel}) we get, formally, a different correlator.
However, the expressions of Eqs. (\ref{B2result}) and (\ref{G2result}) are equivalent and only differ in the contribution of the phase space. 
Furthermore these differences are immaterial when estimating the degree of second-order coherence in the large-scale limit (see 
section \ref{sec5}).

\subsection{Continuity of the effective horizon}
For a reliable implementation of the large-scale limit of the degrees of quantum coherence, a continuous evolution 
of the extrinsic curvature, of the susceptibility and of the effective horizon is mandatory.
For this purpose we shall consider the following expressions for the scale factors across the inflationary transition\footnote{Note that the 
$\gamma$ appearing in Eq. (\ref{contex}) has nothing to do with the $\gamma_{p}$ appearing in Eqs. (\ref{four})--(\ref{AA24a}). This remark avoids potential 
confusions.}:
\begin{eqnarray}
a_{inf}(\tau) &=& \biggl(- \frac{\tau}{\tau_{i}}\biggr)^{- \gamma}, \qquad \tau < - \tau_i,
\nonumber\\
a_{rad}(\tau) &=& \frac{\gamma \tau + (\gamma + 1)\tau_i}{\tau_i}, \qquad \tau \geq - \tau_i,
\label{contex}
\end{eqnarray}
where $\gamma=1$ in the case of an exact de Sitter phase. During a quasi-de Sitter phase, 
the connection between the  conformal time coordinate and  the Hubble rate is given by ${\mathcal H} = aH = -1/[(1-\epsilon)\tau]$ (assuming constant 
slow-roll parameters). According to Eq. (\ref{contex}) the scale factors and  their first time derivatives are continuous, 
i.e.  $a_{rad}(-\tau_{i}) = a_{i}(-\tau_{i})$ and $a^{\prime}_{rad}(-\tau_{i}) = a^{\prime}_{inf}(- \tau_{i})$; therefore 
the extrinsic curvature ${\mathcal H}/a$ is also continuous since ${\mathcal H}_{rad}(-\tau_{i}) = {\mathcal H}_{inf}(- \tau_{i})$. 

The continuous evolution of $\chi$ can then be parametrized in two complementary ways. In the first 
case the susceptibility approaches  exponentially the constant asymptote and the evolution 
of $\chi(\tau)$ across the boundary $\tau = - \tau_{i}$ will then be parametrized as\footnote{If the solution (\ref{A}) is simply matched to 
a constant value of $\chi$ for $\tau> - \tau_{i}$ the first derivative will be discontinuous while the second derivative of $\chi$ at the transition will be
singular. All the parametrizations must then contain a transition regime (as in Eqs. (\ref{A})--(\ref{B}) and (\ref{A1})--(\ref{B1})) which can be studied, though,
 in the sudden limit (i.e., respectively, for $\beta\gg 1$ and $\alpha\gg 1$).}:
\begin{eqnarray}
\chi_{inf}(\tau) &=& \chi_{i} \biggl(- \frac{\tau}{\tau_{i}} \biggr)^{1/2 -\nu}, \qquad \tau< - \tau_{i},
\label{A}\\
\chi_{rad}(\tau) &=& \chi_{i} \biggl[ \biggl(1 - \frac{1 - 2\nu}{2 \beta}\biggr) + \frac{1- 2\nu}{2\beta} e^{- \beta(\tau/\tau_{i} +1)} \biggr], \qquad \tau \geq - \tau_{i}.
\label{B}
\end{eqnarray}
From the explicit expressions of Eqs. (\ref{A}) and (\ref{B}) we have that $\chi_{i}(-\tau_{i}) = \chi_{r}(-\tau_{i})$ and, similarly, 
$\chi_{i}^{\prime}(-\tau_{i}) = \chi_{r}^{\prime}(-\tau_{i})$ implying that both the functions and their first derivatives are continuous. 
The continuity of the susceptibility and of its first derivative implies the continuity of ${\mathcal F}  = \chi^{\prime}/\chi$.
In the cosmic time parametrization we shall have that ${\mathcal F} = a F$ where $F =\dot{\chi}/\chi$ and 
the overdot denotes a derivation with respect to the cosmic time coordinate $t$.
In Eq. (\ref{B}) the rate with which the constant value $\chi_{1}$ is approached is controlled by $\beta$. The interesting physical limit will 
be the one where $\beta \gg 1$: in this limit the transition is continuous but it occurs suddenly. 

The same sudden limit can be studied using a power-law parametrization for the transition regime, like, for instance:
\begin{eqnarray}
\chi_{inf}(\tau) &=& \chi_{i} \biggl(- \frac{\tau}{\tau_{i}} \biggr)^{1/2 -\nu}, \qquad \tau< - \tau_{i},
\label{A1}\\
\chi_{rad}(\tau) &=& \chi_{i} \biggl[ \frac{2(\alpha + \nu) -1}{2\alpha} + \frac{1 - 2 \nu}{2\alpha} \biggl( \frac{\tau}{\tau_{i}} + 1\biggr)^{-\alpha}\biggr], \qquad 
\tau \geq - \tau_{i}.
\label{B1}
\end{eqnarray}
In Eq. (\ref{B1}) the parameter $\alpha\geq 1$ plays the same role of $\beta$ in Eqs. (\ref{A}) and (\ref{B}): it controls the rate of the transition in the intermediate 
regime and as $\alpha$ increases the transition gets more sudden. The expressions 
of Eqs. (\ref{A1}) and (\ref{B1}) are continuous and differentiable, as it can b explicitly checked i.e. $\chi_{inf}(-\tau_{i}) = \chi_{rad}(-\tau_{i})$ and
$\chi_{inf}^{\prime}(-\tau_{i}) = \chi_{rad}^{\prime}(-\tau_{i})$. In spite of the different analytical details, the parametrizations
of Eqs. (\ref{A})--(\ref{B}) and (\ref{A1})--(\ref{B1}) lead to the same results in the sudden limit.
In numerical studies of the problem (see e.g. third paper of \cite{DT2}) the continuous evolution of the susceptibility and of the effective 
horizon have been always enforced even if there are some who 
confuse the sudden approximation (i.e. the regime $\beta \gg 1$ or $\alpha \gg 1$) with a discontinuity of the effective horizon.

\subsection{Evolution of the squeezing parameters} 

According to Eqs. (\ref{four}) and (\ref{AA24})--(\ref{AA24b}) the evolution $r_{p}$, $\gamma_{p}$ and $\alpha_{p}$ 
follows directly from  $u_{p}$ and $v_{p}$:  Eqs. (\ref{AA24})--(\ref{AA24b}) have been derived from Eq. (\ref{five}) 
by means of Eq. (\ref{four}). However, instead of solving  Eqs. (\ref{AA24})--(\ref{AA24b}) it is more practical to 
derive $u_{p}(\tau)$ and $v_{p}(\tau)$,  rephrase the result in terms of the squeezing parameters and take, when needed, the 
large-scale limit. In this procedure Eq. (\ref{five}) and Eqs. (\ref{AA24})--(\ref{AA24b}) can be
used interchangeably in order to simply some of the asymptotic expressions.

When $\tau < -\tau_{i}$, Eq. (\ref{A}) (or Eq. (\ref{A1})) can be inserted into Eq. (\ref{five}) 
and the corresponding solutions will be given by\footnote{In Eq. (\ref{UVinf}) we used the following notation 
$U_{k}(\tau) = u_{k}^{(inf)}(\tau)$ and $V_{k}(\tau) = v_{k}^{(inf)}(\tau)$ to avoid potential confusions with other superscripts.}: 
\begin{eqnarray}
U_{k}(\tau) &=& \frac{{\mathcal N}}{2} \sqrt{- k \tau} \biggl[ H^{(1)}_{\nu}(- k \tau) - i H^{(1)}_{\nu - 1}(- k \tau)\biggr],
\nonumber\\
V_{k}(\tau) &=&  \frac{{\mathcal N}^{*}}{2} \sqrt{- k \tau} \biggl[  i H^{(2)}_{\nu - 1}(- k \tau) - H^{(2)}_{\nu}(- k \tau) \biggr], 
\label{UVinf}
\end{eqnarray}
where ${\mathcal N} = e^{i \pi(\nu +1/2)/2} \sqrt{\pi/2}$; $H^{(1)}_{\nu}(z)= H^{(2)\,\,*}_{\nu}(z)$ are the Hankel functions 
\cite{abr}.   The solution (\ref{UVinf}) is correctly normalized and, as it can be 
explicitly checked $|U_{k}(\tau)|^2 - |V_{k}(\tau)|^2 =1$.

The same strategy leading to Eq. (\ref{UVinf}) could also be 
employed in the regime $\tau > - \tau_{i}$;  the idea would be to insert 
Eqs. (\ref{B}) (or (\ref{B1})) inside Eq. (\ref{five}) and then deduce the corresponding solutions. 
However, if  $\chi$ scales with $(\tau/\tau_{i})$ (i.e. $\chi= \chi(z)$ with $z= \tau/\tau_{i}$)
 the equation for  $(u_{k} + v_{k}^{*})$ obeys, in spite of the functional form of $\chi(z)$
\begin{equation}
\frac{d^2}{d z^2} (u_{k} + v_{k}^{*}) + \biggl[ k^2 \tau_{i}^2 - \frac{1}{\chi} \frac{d^2 \chi}{dz^2}\biggr] (u_{k} + v_{k}^{*}) =0, 
\label{uv}
\end{equation}
where $z = \tau/\tau_{i}$ is the scaling variable. Provided the transition occurs through 
a scaling period where $\chi= \chi(\tau/\tau_{i})$, the first term inside the square bracket of Eq. (\ref{uv}) is always negligible:
$k\tau_{i}$ is at most of order $1$ since the largest amplified wavenumber is ${\mathcal O}(1/\tau_{i})$. In similar terms we also have 
\begin{equation}
 \frac{d^2}{d z^2} (u_{k} - v_{k}^{*}) + \biggl[ k^2 \tau_{i}^2 - \chi \frac{d^2}{dz^2}\biggl(\frac{1}{\chi}\biggr)\biggr] (u_{k} - v_{k}^{*}) =0.
 \label{uv2}
 \end{equation}
The solution of Eqs. (\ref{uv}) and (\ref{uv2}) to lowest order in $k^2 \tau_{i}^2$ can be written as
\footnote{Equations (\ref{uv3}) and (\ref{uv4}) hold under the condition $k \tau_{i} \leq 1$ which is is verified for all the amplified 
modes of the spectrum; this condition is less stringent than 
the usual requirement that the modes are larger than the effective horizon (i.e. $k \tau < 1$).}:
\begin{eqnarray}
&& (u_{k} + v_{k}^{*}) = c_{+}(k) \chi(\tau) + c_{-}(k) \chi \int^{z} \frac{ d z_{1}}{\chi^2(z_{1})},
\label{uv3}\\
&& (u_{k} - v_{k}^{*}) = \frac{\widetilde{c}_{+}(k)}{\chi(\tau)} + \frac{\widetilde{c}_{-}(k)}{\chi(\tau)} \int^{z}  d z_{1} \chi^2(z_{1}).
\label{uv4}
\end{eqnarray}

For an analytically tractable solution it is practical to use an explicit profile such as the one 
of Eq. (\ref{B}). The full solution for $\tau > - \tau_{i}$ is therefore given by\footnote{While this solution 
holds in the case of 
the profile (\ref{A})--(\ref{B}) a similar result 
can be obtained in the case of Eqs. (\ref{A1})--(\ref{B1}) 
but, for the sake of conciseness, the details will be skipped. }:
\begin{eqnarray}
u_{k}(x_{i}, \tau) &=&\frac{{\mathcal N}\, \sqrt{x_{i}}}{ 2 C^2 \beta} \frac{ e^{- \beta(\tau/\tau_{i} +1)}}{D + C e^{\beta(\tau/\tau_{i} +1)}}\biggl\{C^2 \biggl[D + C e^{\beta(\tau/\tau_{i} +1)}\biggr]^2\beta H_{\nu}^{(1)}(x_{i}) +  
\nonumber\\
&+& H^{(1)}_{\nu-1}(x_{i}) \biggl[D^2 (D -1) x_{i} + C D( 2 D -1) e^{ 2\beta(\tau/\tau_{i} +1)}( D x_{i} - i \beta) 
\nonumber\\
&-& \biggl[ D + C e^{\beta(\tau/\tau_{i} +1)}\biggr]^2 x_{i} \ln{ \biggl( D + C e^{\beta(\tau/\tau_{i} +1)}}\biggr)\biggr]\biggr\},
\label{uex}\\
v_{k}(x_{i}, \tau) &=&\frac{{\mathcal N}^{*}\, \sqrt{x_{i}}}{ 2 C^2 \beta} \frac{ e^{- \beta(\tau/\tau_{i} +1)}}{D + C e^{\beta(\tau/\tau_{i} +1)}}\biggl\{-C^2 \biggl[D + C e^{\beta(\tau/\tau_{i} +1)}\biggr]^2\beta H_{\nu}^{(2)}(x_{i}) +  
\nonumber\\
&-& H^{(2)}_{\nu-1}(x_{i}) \biggl[D^2 (D -1) x_{i} + C D( 2 D -1) e^{ 2\beta(\tau/\tau_{i} +1)}( D x_{i} - i \beta) 
\nonumber\\
&-& \biggl[ D + C e^{\beta(\tau/\tau_{i} +1)}\biggr]^2 x_{i} \ln{ \biggl( D + C e^{\beta(\tau/\tau_{i} +1)}}\biggr)\biggr]\biggr\},
\label{vex}
\end{eqnarray}
where, for simplicity, we defined $C= 1 - (1- 2\nu)/(2 \beta)$ and $D =  (1- 2\nu)/(2 \beta)$;
for $\tau\geq -\tau_{i}$ the solutions $u^{(rad)}_{k}(\tau)$ and $v^{(rad)}_{k}(\tau)$  have been denoted, respectively, by $u_{k}(\tau)$ and $v_{k}(\tau)$.
It follows from Eqs. (\ref{uex}) and (\ref{vex}) that $|u_{k}(x_{i},\tau)|^2 - |v_{k}(x_{i},\tau)|^2=1$. Note that the obtained 
solution, as required, is continuous and differentiable everywhere and, in particular, at the transition point $\tau= - \tau_{i}$ (recall, for this 
purpose, that $C+ D =1$). 

\subsection{Crossing of the effective horizon}
The condition defining the time when a given mode reenters the effective horizon is obtained by requiring $\chi_{rad}^{\prime\prime}/\chi_{rad} = k^2$;
the latter condition implies: 
\begin{equation}
\frac{\tau_{re}}{\tau_{i}} + 1 = \frac{1}{\beta} \ln{\biggl[ \frac{ D (\beta^2 - x_{i}^2)}{C x_{i}^2} \biggr]},
\label{exact1}
\end{equation}
where Eq. (\ref{B}) has been explicitly used. 
Equation (\ref{exact1}) defines the crossing of the effective horizon as a function of $x_{i} = k \tau_{i}$. Since $k \tau_{i} \leq 1$ ($k \tau_{i} \ll 1$ for the typical 
scale of the gravitational collapse) we will have that 
\begin{equation}
\frac{\tau_{re}}{\tau_{i}} = - \frac{2}{\beta}\ln{\biggl(\frac{k}{a_{i} H_{i}}\biggr)} - \frac{1}{\beta}\ln{\biggl|\frac{C}{D}\biggr|} -\frac{x_i^2}{\beta^3} 
+ {\mathcal O} (x_{i}^4),
\label{tre}
\end{equation}
where  $k \tau_{i} = k/(a_{i} H_{i})$.
To get an idea of the accuracy of this expansion we can compute $k/(a_{i} H_{i})$ in terms of the fiducial parameters 
of the concordance scenario:
\begin{equation}
\frac{k}{a_{i} H_{i}} = 3.71\times 10^{-24} \, \biggl(\frac{k}{\mathrm{Mpc}^{-1}}\biggr) \, \biggl(\frac{\epsilon}{0.01}\biggr)^{-1/4} \, 
\biggl(\frac{{\mathcal A}_{\mathcal R}}{2.41\times 10^{-9}} \biggr)^{-1/4},
\label{ex0}
\end{equation}
where ${\mathcal A}_{{\mathcal R}}$ is the amplitude of the power spectrum of scalar fluctuations 
at the pivot scale $k_{p} =0.002\, \mathrm{Mpc}^{-1}$.

To compute the degrees of quantum coherence we must fix a reference time and we shall take this 
reference time to coincide with $\tau_{re}$. Alternatively one can keep the time-scale 
generic and expand the relevant correlation functions in the limit $x_{i} \ll 1$.
Inserting then Eq. (\ref{exact1}) into Eq. (\ref{uex}) and (\ref{vex}) we obtain\footnote{It can 
be explicitly verified that $|u_{k}(x_{i}, \nu,\beta)|^2 - |v_{k}(x_{i}, \nu,\beta)|^2 =1$, as required 
by the commutation relations. }
\begin{eqnarray}
u_{k}(x_{i}, \nu,\beta) &=& \frac{{\mathcal N}}{2 C\beta^2 (\beta^2 - x_{i}^2)} \sqrt{x_i} \biggl\{ C^2 \beta^4 H^{(1)}_{\nu}(x_{i})
-  H^{(1)}_{\nu-1}(x_{i}) \biggl[ i  (x_i^4 - i x_i^3 \beta - 2 x_i^2 \beta^2 
\nonumber\\
&+&   i D x_i \beta^3 + \beta^4) +x_i \beta^3 \log{\biggl(\frac{D \beta^2}{x_{i}^2}\biggr)}\biggr]\biggr\},
\label{uexpand1}\\
v_{k}(x_{i}, \nu,\beta) &=& \frac{{\mathcal N}^{*}}{2 C\beta^2 (\beta^2 - x_{i}^2)} \sqrt{xi} \biggl\{-C^2 \beta^4 H^{(2)}_{\nu}(x_{i}) 
+ H^{(2)}_{\nu-1}(x_{i}) \biggl[  i  (x_i^4 - i x_i^3 \beta - 2 x_i^2 \beta^2
\nonumber\\
&+&   i D x_i \beta^3 + \beta^4) + x_i \beta^3 \log{\biggl(\frac{D \beta^2}{x_{i}^2}\biggr)}\biggr]\biggr\}.
\label{vexpand2}
\end{eqnarray}
These equations are still exact but they can be expanded around the effective horizon. From Eqs. (\ref{uexpand1}) and (\ref{vexpand2})  the squeezing parameters 
can be obtained, as we shall now show. Since, by definition $\overline{n}_{k}(x_{i}, \nu, \beta) = |v_{k}(x_{i},\nu,\beta)|^2$ 
 the average multiplicity can be computed by expanding, at once, the whole expression:
\begin{eqnarray}
|v_{k}(x_{i},\nu,\beta)|^2 &=&  \biggl(\frac{x_{i}}{2}\biggr)^{- 2 \nu} \biggl[ \frac{C^2\, \Gamma^2(\nu)}{8 \pi} x_{i} + {\mathcal O}(x_{i}^3)\biggr] 
\nonumber\\
&+& \biggl\{ - \frac{1}{2} + \frac{1}{4 \tan{\nu\pi}} \bigg[ - \frac{1}{C^2 (\nu -1)} - \frac{C^2}{\nu} - \frac{2 D}{\beta} + \frac{2}{\beta} \ln{\biggl(\frac{D \beta^2}{x_{i}^2}\biggr)}\biggr] x_{i} + {\mathcal O}(x_{i}^2)\biggr\}
\nonumber\\
&+& \biggl(\frac{x_{i}}{2}\biggr)^{2 \nu} \biggl[ \frac{\pi}{2 C^2 \Gamma^2(\nu) x_{i} \sin^2{\nu\pi}} + {\mathcal O}(x_{i})\biggr]
\label{vsq}
 \end{eqnarray}
where we extensively used that $C+ D = 1$ (and hence that $C^2 - D^2 = C- D$).  The same result of Eq. (\ref{vsq}) can be obtained if we
expand around the effective horizon but keep the Hankel functions in their exact form. The result of this procedure is:
\begin{eqnarray}
\overline{n}_{k}(x_{i}, \nu, \beta) &=& \frac{\pi \, x_{i}}{8} \biggl[ \frac{C^2}{4} \, |H_{\nu}^{(1)}(x_{i})|^2 + \frac{1}{C^2} \, |H_{\nu-1}^{(1)}(x_{i})|^2 -1\biggr]
\nonumber\\
&+& \frac{\pi x_{i}^2}{8\beta}\biggl[ H_{\nu}^{(1)}(x_{i}) H_{\nu-1}^{(2)}(x_{i}) + H_{\nu-1}^{(1)}(x_{i}) H_{\nu}^{(2)}(x_{i})\biggr] \biggl(D + 2 \ln{x_{i}} - \ln{D \beta^2}\biggr)
\nonumber\\
&+& {\mathcal O}(x_{i}^3).
\label{vsq2}
\end{eqnarray}
Recalling that $e^{- i \varphi_{k}(\tau)} |u_{k}(\tau)| = u_{k}(\tau)$
we can express $\varphi_{k}(\tau_{re})$ in a closed form:
\begin{equation}
e^{- i \varphi_{k}(\tau_{re})} = e^{i (\nu + 1/2)\pi/2} \biggl[ - i  - \frac{x_{i}}{2 C^2 (\nu -1) \beta^2}+ {\mathcal O}(x_{i}^2) \biggr]
\label{phiex}
\end{equation}
Equation (\ref{phiex}) can be obtained by writing $u_{k}(x_{i}, \nu, \beta)$ as
\begin{equation}
u_{k}(x_{i}, \nu, \beta) = e^{i (\nu + 1/2)\pi/2}\biggl[ Q_{1}(x_{i}, \nu, \beta) + i \,   P_{1}(x_{i}, \nu, \beta)  \biggr],
\end{equation}
where $ Q(x_{i}, \nu, \beta) $ and $P(x_{i}, \nu, \beta)$ are both real and given by:
\begin{eqnarray}
Q_{1}(x_{i}, \nu, \beta) &=&\frac{ \sqrt{\pi x_{i}}}{2^{3/2} C \beta^2 (\beta^2 - x_{i}^2)}\biggl[ C^2 \beta^4 J_{\nu}(x_{i}) + (x_{i}^2 - \beta^2)^2 Y_{\nu -1}(x_{i}) 
\nonumber\\
&-& x_{i} \beta J_{\nu-1}(x_{i}) ( x_{i}^2 - D \beta^2) + \beta^2 \ln{D\beta^2} - 2 \beta^2 \ln{x_{i}}\biggr]
\nonumber\\
P_{1}(x_{i}, \nu, \beta) &=& -  \frac{ \sqrt{\pi x_{i}}}{2^{3/2} C \beta^2 (\beta^2 - x_{i}^2)} \biggl[ (x_{i}^2 - \beta^2)^2  J_{\nu-1}(x_{i}) - C^2 \beta^2 Y_{\nu }(x_{i}) 
\nonumber\\
&+& x_{i} \beta Y_{\nu-1}(x_{i}) ( x_{i}^2 - D \beta^2 + \beta^2 \ln{D\beta^2} - 2 \beta^2 \ln{x_{i}})\biggr].
\end{eqnarray}
Exactly with the same strategy we can compute $\gamma_{k}$ which is given by 
\begin{equation}
e^{- i[\varphi_{k}(\tau_{re})- \gamma_{k}(\tau_{re})]} =\frac{v_{k}(x_{i}, \nu,\beta)}{ |v_{k}(x_{i}, \nu,\beta)|} = e^{- i (\nu + 1/2)\pi/2}\biggl[- i - \frac{x_{i}}{2 C^2 (\nu -1)} \biggr].
\label{secph}
\end{equation}
By combining Eqs. (\ref{phiex}) and (\ref{secph}) we also have that  
\begin{equation}
e^{- i \alpha_{k}(\tau_{re})} = - 1 + \frac{i x_{i}}{2 C^2 (\nu -1)} \biggl( 1 + \frac{1}{\beta^2}\biggr) + {\mathcal O}(x_{i}^2).
\end{equation}
With the results obtained so far we shall be able to discuss in detail the degrees of first-order and second-order 
coherence. 

\renewcommand{\theequation}{5.\arabic{equation}}
\setcounter{equation}{0}
\section{Degrees of coherence in the large-scale limit}
\label{sec5}
The degrees of first-order and second-order coherence will now be computed. We shall then contrast the 
results with the benchmark values obtained in the context of the single-mode 
approximation.
\subsection{First-order coherence}
From the discussion of section \ref{sec4}, the degrees of coherence can be computed at any time 
$\tau_{i} < \tau \leq \tau_{re}$ but the most relevant reference time is $\tau = {\mathcal O}(\tau_{re})$;
in this case, $v_{k}(\tau)$ and $u_{k}(\tau)$ are given by Eqs. (\ref{uex}) and (\ref{vex}).
From Eqs. (\ref{firstEB}) and (\ref{trace}), after angular integration, the first-order correlation function at separate space-time points are 
\begin{eqnarray}
{\mathcal B}^{(1)}(x_{1},\, x_{2}) &=& {\mathcal E}^{(1)}(x_{1}, x_{2}) 
\nonumber\\
&=& \frac{C^2}{64 \pi^2} \int d p \, p^{3}\, \sqrt{p\, \tau_{2}}\, H^{(2)}_{\nu}(p\tau_{2})\,\, \sqrt{p \tau_{1}}\,H^{(1)}_{\nu}(p\tau_{1})\, 
j_{0}(p r),
\label{CC3b}\\
{\mathcal G}^{(1)}(x_{1},\, x_{2}) &=& \frac{C^2}{64 \pi^2} \int p\,d p\, \sqrt{p\, \tau_{2}}\, H^{(2)}_{\nu}(p\tau_{2})\,\, \sqrt{p \tau_{1}}\,H^{(1)}_{\nu}(p\tau_{1})\, j_{0}(p r),
\label{CC3c}
\end{eqnarray}
where $j_{0}(k_{1} r)$ denotes the spherical Bessel function of zeroth order \cite{abr}.
From Eqs. (\ref{CC3b})--(\ref{CC3c}) the normalized degree of first-order coherence defined
 in Eqs. (\ref{corEB}) and (\ref{corG}) becomes:
\begin{eqnarray}
g_{\mathcal B}^{(1)}(\vec{x}_{1},\, \vec{x}_{2};\, \tau_{1},\, \tau_{2}) &=& 
g_{\mathcal E}^{(1)}(\vec{x}_{1},\, \vec{x}_{2};\, \tau_{1},\, \tau_{2})
\nonumber\\
&=& \frac{\int d p_{1} \, p_{1}^3\, v_{p_{1}}^{*}(\tau_{1})\, v_{p_{1}}(\tau_{2}) \, j_{0}(p_{1} r) }{\sqrt{\int d p_{1} \,p_{1}^3\,  v_{p_{1}}^{*}(\tau_{1})\, v_{p_{1}}(\tau_{2})} \sqrt{\int d p_{2} \, p_{2}^3\, v_{p_{2}}^{*}(\tau_{2})\, v_{p_{2}}(\tau_{2})}},
\label{CC4}\\
g_{\mathcal G}^{(1)}(\vec{x}_{1},\, \vec{x}_{2};\, \tau_{1},\, \tau_{2})&=& \frac{\int d p_{1} \,p_{1}\, v_{p_{1}}^{*}(\tau_{1})\,
 v_{p_{1}}(\tau_{2}) \, j_{0}(p_{1}r)}{\sqrt{\int d p_{1} p_{1}\, 
 v_{p_{1}}^{*}(\tau_{1})\, v_{p_{1}}(\tau_{2})} \sqrt{\int d p_{2} \, 
 p_{2}\, v_{p_{2}}^{*}(\tau_{2})\, v_{p_{2}}(\tau_{2})}}.
\label{CC4a}
\end{eqnarray}
Using Eqs. (\ref{CC3b}) and (\ref{CC3c}) into Eq. (\ref{CC4a}) 
the numerators and the denominators of Eq. (\ref{CC4}) 
depend on $\tau_{1}$ and $\tau_{2}$ but, as a consequence of Eqs. (\ref{uex}) and (\ref{vex}),
 this dependence simplifies when computing the degrees 
of quantum coherence in the large-scale limit. Therefore the final form of Eqs. (\ref{CC4})--(\ref{CC4a}) can be written as:
\begin{eqnarray}
g_{\mathcal B}^{(1)}(r) &=& g_{\mathcal E}^{(1)}(r) = \frac{\int d p\, p^{ 5 - 2 \nu} \, j_{0}(p r)}{\int d p\, p^{ 5 - 2 \nu}} \to 1,
\label{CC5a}\\
g_{\mathcal G}^{(1)}(r) &=&  \frac{\int d p\, p^{ 3 - 2 \nu} \, j_{0}(p r)}{\int d p\, p^{ 3 - 2 \nu}} \to 1,
\label{CC5b}
\end{eqnarray}
where the integrals are evaluated over all the modes larger than the effective Hubble radius and the second relation clearly holds in the 
limit $k_{1} r \ll 1$ (corresponding to large angular separations). Equations (\ref{CC5a}) and (\ref{CC5b}) remain clearly valid in the 
zero time-delay limit (i.e. $\tau_{1}\to \tau_{2}$).

\subsection{Second-order coherence}
According to Eq. (\ref{ident}) the space-time points can be identified two by two and, in this case, Eqs. (\ref{B2result}) and (\ref{G2result}) define 
the intensity correlation which is typical of the HBT effect. More specifically, when $x_{3}\to x_{1}$ and $x_{4} \to x_{2}$ 
Eqs. (\ref{B2result}) and (\ref{G2result}) become:
\begin{eqnarray}
{\mathcal B}^{(2)}(x_{1},\,x_{2}) &=&\int \frac{d^{3} p_{1}}{(2\pi)^3} \, p_{1}\,  \int \frac{d^{3} p_{2}}{(2\pi)^3} \, p_{2} \biggl\{ |v_{p_{1}}(\tau_{1})|^2 
|v_{p_{2}}(\tau_{2})|^2
\nonumber\\
&+& \frac{1}{4} \biggl[ 1 +  \frac{(\vec{p}_{1}\cdot\vec{p}_{2})^2}{p_{1}^2 p_{2}^2}\biggr]\biggl[ v_{p_{1}}^{*}(\tau_{1}) v_{p_{2}}^{*}(\tau_{2}) v_{p_{2}}(\tau_{1}) v_{p_{1}}(\tau_{2}) e^{- i (\vec{p}_{1} - \vec{p}_2) \cdot \vec{r}}
\nonumber\\
&+& v_{p_{1}}^{*}(\tau_{1}) u_{p_{1}}^{*}(\tau_{2}) u_{p_{2}}(\tau_{1}) v_{p_{2}}(\tau_{2}) e^{- i (\vec{p}_{1} + \vec{p}_{2})\cdot\vec{r}}\biggr]\biggr\}.
\end{eqnarray}
The normalized degrees of second-order coherence are 
\begin{eqnarray}
g_{{\mathcal E}}^{(2)}(x_{1},\, x_{2}) &=& \frac{{\mathcal E}^{(2)}(x_{1}, \, x_{2})}{{\mathcal E}^{(1)}(x_{1}, \, x_{1}) \, {\mathcal E}^{(1)}(x_{2}, \, x_{2})},
\nonumber\\
g_{{\mathcal B}}^{(2)}(x_{1},\, x_{2}) &=& \frac{{\mathcal B}^{(2)}(x_{1}, \, x_{2})}{{\mathcal B}^{(1)}(x_{1}, \, x_{1}) \, {\mathcal B}^{(1)}(x_{2}, \, x_{2})},
\label{seconEB}\\ 
g_{{\mathcal G}}^{(2)}(x_{1},\, x_{2}) &=& \frac{{\mathcal G}^{(2)}(x_{1}, \, x_{2})}{{\mathcal G}^{(1)}(x_{1}, \, x_{1}) \, {\mathcal G}^{(1)}(x_{2}, \, x_{2})}.
\label{seconG}
\end{eqnarray}
Equations (\ref{seconEB}) and (\ref{seconG}) are nothing but
the correlations of the intensity. Up to terms that are small in the large-scale 
limit, ${\mathcal B}^{(2)}(x_{1}, \, x_{2})$ and ${\mathcal E}^{(2)}(x_{1}, \, x_{2})$ can  be expressed as 
\begin{eqnarray}
{\mathcal B}^{(2)}(x_{1},\,x_{2}) &=& {\mathcal E}^{(2)}(x_{1},\,x_{2}) = \int \frac{d^{3} p_{1}}{(2\pi)^3} \, p_{1}\, \overline{n}_{p_{1}}(\tau_{1}) \int \frac{d^{3} p_{2}}{(2\pi)^3} \, p_{2} \, \overline{n}_{p_{2}}(\tau_{2}) 
\nonumber\\
&\times& \biggl\{ 1 + \frac{1}{4} [ 1 +   (\hat{p}_{1} \cdot\hat{p}_{2})^2]\biggl[1+ e^{- i (\vec{p}_{1} - \vec{p}_2) \cdot \vec{r}} 
+ e^{- i (\vec{p}_{1} + \vec{p}_{2}) \cdot\vec{r}}
\nonumber\\
&+& {\mathcal O} (p_{1} p_{2} \tau_{1} \tau_{2}) \biggr]\biggr\}.
\label{EBsec}
\end{eqnarray}
Equation (\ref{EBsec}) follows from the observation that 
\begin{eqnarray}
\frac{ u_{p_{1}}^{*}(\tau_{1}) \, v_{p_{1}}^{*}(\tau_{1})  u_{p_{2}}(\tau_{2}) \, v_{p_{2}}(\tau_{2}) }{\overline{n}_{p_{1}}(\tau_{1}) \, 
\overline{n}_{p_{2}}(\tau_{2})}=  1+ {\mathcal O} (p_{1} p_{2} \tau_{1} \tau_{2}).
\label{CC9a}
\end{eqnarray}
The angular integrals appearing in Eq. (\ref{EBsec}) can be performed by expressing the momenta in polar coordinates and the 
result in terms of the degree of second-order coherence becomes 
\begin{eqnarray}
g^{(2)}_{{\mathcal B}}(r, \tau_{1}, \tau_{2}) &=& g^{(2)}_{{\mathcal E}}(r, \tau_{1}, \tau_{2})
\nonumber\\
&=& \frac{\int p^{3}_{1} \,d p_{1}  \overline{n}_{p_{1}}(\tau_{1}) \,\int p^{3}_{2} \, d p_{2}  \overline{n}_{p_{2}}(\tau_{2})J(r, p_1, p_2)}{\int p^{3}_{1} d p_{1}  \overline{n}_{p_{1}}(\tau_{1}) \int p^{3}_{2} d p_{2}  \overline{n}_{p_{2}}(\tau_{2})}
\end{eqnarray}
where $J(r, p_1, p_2)$ is defined as:
\begin{eqnarray}
J(r, p_1, p_2)&=& 1  + \frac{\cos{p_1 r} (3 p_2 r \cos{p_2 r} + ( p_2^2 r^2 - 3) \sin{p_2 r}) }{ p_1^2 p_2^3 r^5 }
\nonumber\\
&+& \frac{\sin{p_1 r} [p_2 r (p_1^2 r^2 - 3 ) \cos{p_2 r} 
+ (3 - p_2^2 r^2 + p_1^2 r^2 (p_2^2 r^2 -1 )) \sin{p_2 r}]}{ p_1^3 p_2^3 r^6 }
\nonumber\\
&=& \frac{5}{3} - \frac{r^2 (p_{1}^2+ p_{2}^2)}{9} + \frac{(5 p_{1}^4 +18 p_{1}^2 p_{2}^2 + 5 p_{2}^4) r^4}{900} + {\mathcal O}( p^5 r^5).
\label{JEQ}
\end{eqnarray}
The last line of Eq. (\ref{JEQ}) corresponds to the large-scale limit obtained by expanding the exact expression for 
$p_{1} r < 1$ and $p_{2} r < 1$; note that $p$ in the correction denotes a generic momentum.
If applied to ${\mathcal G}^{(2)}(x_{1}, x_{2})$ the same analysis  leads to the following expression for the 
second-order coherence:
\begin{eqnarray}
g^{(2)}_{{\mathcal G}}(r, \tau_{1}, \tau_{2}) = \frac{\int p_{1} d p_{1}  \overline{n}_{p_{1}}(\tau_{1}) \int p_{2} d p_{2}  \overline{n}_{p_{2}}(\tau_{2}) J(r, p_1, p_2)}{\int p_{1} d p_{1}  \overline{n}_{p_{1}}(\tau_{1}) \int p_{2} d p_{2}  \overline{n}_{p_{2}}(\tau_{2})}.
\end{eqnarray}
The large-scale limit of the degree of second-order coherence can then be written as 
\begin{equation}
\lim_{\tau_{1} \to \tau_{2}} g_{{\mathcal B}}^{(2)}(r,\tau_{1}, \tau_{2}) = \lim_{\tau_{1} \to \tau_{2}}  g_{{\mathcal E}}^{(2)}(r,\tau_{1}, \tau_{2}) = \lim_{\tau_{1} \to \tau_{2}}  g_{{\mathcal G}}^{(2)}(r,\tau_{1}, \tau_{2}) \to \frac{5}{3}.
\label{CC11}
\end{equation}
The result of Eq. (\ref{CC11}) holds  in the zero time-delay limit 
$\tau_{1}- \tau_{2}=0$.

\subsection{Physical interpretation}
Equations (\ref{CC5a})--(\ref{CC5b}) and (\ref{CC11}) differ from the ones obtainable
in the conventional single-mode approximation which is often mentioned in quantum optical applications. In short we could say that 
while the degree of second-order coherence should go to $3$ for a squeezed state, we got $5/3$ (see Eq. (\ref{CC11})).
The rationale for the disagreement, as we shall see hereunder, has to do with the polarizations. 

More specifically, according to the results of Eqs. (\ref{CC5a})--(\ref{CC5b}) and (\ref{CC11})  in the zero time-delay limit (i.e. $(\tau_{1} -\tau_{2}) \to 0$) and for large-scales, the degrees quantum states are first-order coherent
(i.e. $g_{{\mathcal B}}^{(1)}(0) = g_{{\mathcal E}}^{(1)}(0)= g_{{\mathcal G}}^{(1)}(0) =1$) 
but not second-order coherent (i.e. $g_{{\mathcal B}}^{(2)}(0) = g_{{\mathcal E}}^{(2)}(0)= g_{{\mathcal G}}^{(2)}(0) = 5/3$).
To facilitate the comparison with the forthcoming considerations we denoted by $g^{(1)}_{X}(0)$ and $g^{(2)}_{X}(0)$
(with $X = {\mathcal B},\,{\mathcal E},\,{\mathcal G}$) the first- and second-order degrees of quantum coherence in the zero time-delay limit. 

In quantum optics the numerical values of the degrees of first- and second-order coherence 
are customarily classified by considering a single mode of the field and a single 
polarization\footnote{This approximation is often referred to as single mode 
quantum optics (see, e.g. chapter 5 of Ref. \cite{loudonb} and also \cite{mandel}).
The rationale for this approximation is that many experiments use plane parallel light beams whose transverse
 intensity profiles are not important for the measured quantities. As a consequence it is often sufficient in interpreting 
 the data to consider the light beams as exciting a single mode of the field. In actual interferometry the electric field is first split into two components 
through the beam splitter, then it is time-delayed and finally recombined at the correlator. 
The limit of zero time delay between the signals is commonly used, in both cases,  to characterize the statistical properties of the source.}. 
For a single mode of the field the degrees of first- and second-order coherence are defined as:
\begin{eqnarray}
\overline{g}^{(1)}(\tau_{1}, \tau_{2}) &=& \frac{\langle \hat{a}^{\dagger}(\tau_{1}) \, \hat{a}(\tau_{2})\rangle}{\sqrt{\langle \hat{a}^{\dagger}(\tau_{1}) \, \hat{a}(\tau_{1})\rangle}
\, \sqrt{\langle \hat{a}^{\dagger}(\tau_{2}) \, \hat{a}(\tau_{2})\rangle}},
\label{g1qm}\\
\overline{g}^{(2)}(\tau_{1}, \tau_{2}) &=& \frac{\langle \hat{a}^{\dagger}(\tau_{1})  \hat{a}^{\dagger}(\tau_{2}) \, \hat{a}(\tau_{2})\, \hat{a}(\tau_{1})\rangle}{\langle \hat{a}(\tau_{1}) \, \hat{a}(\tau_{1})\rangle \langle a^{\dagger}(\tau_{2}) \, a(\tau_{2})\rangle},
\label{g2qm}
\end{eqnarray}
where the overline at the left hand side distinguishes Eqs. (\ref{g1qm}) and (\ref{g2qm}) 
from Eqs. (\ref{seconEB}) and (\ref{seconG}) holding in the general case. 
Equations (\ref{g1qm}) and (\ref{g2qm}) define, respectively, the degrees of first and second-order temporal coherence: in the zero time-delay limit 
$\tau_{1}- \tau_{2}\to 0$ and, in this case, the degree of second-order coherence will be denoted by 
$\overline{g}^{(2)}$. For a single-mode coherent state (i.e. $\hat{a} |\alpha \rangle = \alpha  |\alpha \rangle$), Eqs. (\ref{g1qm}) and (\ref{g2qm}) imply 
\begin{equation}
\overline{g}^{(1)} = \overline{g}^{(2)} =1,
\label{coherent}
\end{equation}
so that a coherent state is both first-order and second-order coherent in the single mode approximation.
For a chaotic state in the single approximation the statistical weights of the the density matrix are provided by the 
Bose-Einstein distribution \cite{mandel,loudonb} and the results for the degrees of coherence imply: 
\begin{equation}
\overline{g}^{(1)}=1, \qquad \overline{g}^{(2)}=2,
\label{chaotic}
\end{equation}
so that the degree of second-order coherence is twice the result of a coherent state.
In the case of a Fock state $\overline{g}^{(2)} = (1 - 1/n )< 1$ showing that Fock states lead always to sub-Poissonian behaviour and they are anti-bunched \cite{mandel,loudonb}.  Let us now come to the most interesting case for the present discussion, namely  the case of a squeezed state \cite{LK},
corresponding\footnote{For simplicity,  
the phases have been fixed to zero since they do not affect the degree of second-order coherence in the 
single-mode approximation. } to $\hat{a} = \cosh{r} \hat{b} - \sinh{r} \hat{b}^{\dagger}$. Taking the limit of zero time-delay and inserting these expressions in Eq. (\ref{g2qm}) we have that: 
\begin{equation}
\overline{g}^{(1)}=1, \qquad\overline{g}^{(2)} = 3 + \frac{1}{\overline{n}}, \qquad \overline{n} = \sinh^2{r}.
\label{squeezed}
\end{equation}
Equation (\ref{squeezed}) also implies that in the limit $\overline{n}\gg 1$ the degree of second-order coherence 
goes to $3$.

In the single-mode approximation, chaotic light is an example of bunched 
 quantum state (i.e. $\overline{g}^{(2)}> 1$ implying more degree of second-order coherence 
 than in the case of a coherent state). Fock states are instead antibunched (i.e. $\overline{g}^{(2)} <1$) implying 
 a degree of second-order coherence smaller than in the case of a coherent state. 
Finally squeezed light is bunched and also superchaotic, meaning that the degree of second-order 
coherence is larger than in the case of thermal state.

Based on the single-mode approximation, we have that the degree of second-order coherence 
of our problem should have implied that $\overline{g}^{(2)}_{X} \to 3 $, for $X = {\mathcal B},\,{\mathcal E},\,{\mathcal G}$.
We instead obtained $\overline{g}^{(2)}_{X} \to 5/3 $ (and $\overline{g}^{(2)}_{X} \to 1 $). 
The reason for this apparent disagreement stems from the contribution 
of the polarizations to the degree of second-order coherence. 

To prove this statement let us consider the case of a scalar field. For this analysis we shall adapt 
the results of Ref. \cite{mg2} valid in the case of the scalar modes of the geometry.  
Recalling the results of Eqs. (\ref{corr1a}), (\ref{S2result})
 the correlation function of Eq. (\ref{scalar1})  ( when $x_{1}= x_{3}$ and $x_{2} = x_{4}$)
describes the interference of two beams with intensities $\hat{{\mathcal I}}(\vec{x}_{1}, \tau_{1})$ and $\hat{{\mathcal I}}(\vec{x}_{2}, \tau_{2})$, i.e. 
\begin{eqnarray}
{\mathcal G}^{(2)}(x_{1},\, x_{2})  &=& \langle \hat{{\mathcal I}}(\vec{x}_{1}, \tau_{1}) \, \hat{{\mathcal I}}(\vec{x}_{2}, \tau_{2}) \rangle =
\frac{1}{4} \int \frac{d^{3} k_{1}}{k_{1} (2\pi)^3} \int \frac{d^{3} k_{2}}{k_{2} (2\pi)^3} 
\nonumber\\
&\times& \biggl\{ |v_{k_{1}}(\tau_{1})|^2 \, |v_{k_{2}}(\tau_{2})|^2 \biggl[ 1 + e^{- i (\vec{k}_{1}- \vec{k}_{2}) \cdot\vec{r}}\biggr]
\nonumber\\
&+& v_{k_{1}}^{*}(\tau_1)\,u_{k_{1}}^{*}(\tau_2) \,u_{k_{2}}(\tau_{1})\, v_{k_{2}}(\tau_{2})\,e^{- i (\vec{k}_{1}+ \vec{k}_{2}) \cdot\vec{r}}\biggr\},
\label{S6a}
\end{eqnarray}
where, as usual, $\vec{r} = \vec{x}_{1} - \vec{x}_{2}$. If we perform the angular integrations, the degree of second-order 
coherence becomes, in this case, 
\begin{eqnarray}
g^{(2)}(\vec{r}, \tau_{1},\tau_{2}) &=& \frac{ \langle \hat{{\mathcal I}}(\vec{x}_{1}, \tau_{1}) \, \hat{{\mathcal I}}(\vec{x}_{2}, \tau_{2})\rangle}{\langle \hat{{\mathcal I}}(\vec{x}_{1}, \tau_{1})\rangle \langle  \hat{{\mathcal I}}(\vec{x}_{2}, \tau_{2}) \rangle} 
\nonumber\\
&=& 1 + \frac{\int k_{1} d k_{1} |v_{k_{1}}(\tau_{1})|^2 \, j_{0}(k_{1} r) \,\,\int k_{2} d k_{2} |v_{k_{2}}(\tau_{2})|^2\, j_{0}(k_{2} r) }{\int k_{1} \,d k_{1} |v_{k_{1}}(\tau_{1})|^2\,\int k_{2} \,d k_{2} |v_{k_{2}}(\tau_{2})|^2}
\nonumber\\
&+&  \frac{\int k_{1} d k_{1} \,u_{k_{1}}^{*}(\tau_{2}) v_{k_1}^{*}(\tau_{1})\, j_{0}(k_{1} r) \,\,\int k_{2} d k_{2} \, u_{k_2}(\tau_{1})v_{k_{2}}(\tau_{2})\,j_{0}(k_{2} r)}{\int k_{1} d k_{1} |v_{k_{1}}(\tau_{1})|^2 \,\,\int k_{2} d k_{2} |v_{k_{2}}(\tau_{2})|^2}.
\label{S8}
\end{eqnarray}
Using now of the same observation of Eq. (\ref{CC9a}) we have that the degree of second-order coherence in the scalar case 
becomes 
\begin{eqnarray}
g^{(2)}(\vec{r},\tau_{1},\tau_{2}) &=& 1 + 2 \frac{\int k_{1} d k_{1} j_{0}(k_{1} \, r)\, \overline{n}_{k_{1}}(\tau_{1}) \, \int k_{2} d k_{2} j_{0}(k_{2} \, r)\, \overline{n}_{k_{2}}(\tau_{2})}{\int k_{1} d k_{1}  \overline{n}_{k_{1}}(\tau_{1}) \, \int k_{2} d k_{2}\, \overline{n}_{k_{2}}(\tau_{2})}
\nonumber\\
&+&  \frac{\int k_{1} d k_{1} j_{0}(k_{1} \, r)/\sqrt{\overline{n}_{k_{1}}(\tau_{1})}\,  \, \int k_{2} d k_{2} j_{0}(k_{2} \, r)/\sqrt{\overline{n}_{k_{2}}(\tau_{2})}\, }{\int k_{1} d k_{1} \overline{n}_{k_{1}}(\tau_{1})\int k_{2} d k_{2}\, \overline{n}_{k_{2}}(\tau_{2})}.
\label{S10}
\end{eqnarray}
The large-scale limit the spherical Bessel functions go to $1$ and therefore 
Eq. (\ref{S10}) becomes: 
\begin{equation}
g^{(2)}(r,\tau_{1}, \tau_{2}) \to  3, \qquad \lim_{\tau_{1} \to \tau_{2}} g^{(2)}(r,\tau_{1}, \tau_{2}) = g^{(2)}(r,\tau).
\label{S11}
\end{equation}
The result of Eq. (\ref{CC11}) holds also in the zero time-delay limit 
$\tau_{1}- \tau_{2}=0$. This analysis demonstrates that the degree of second-order coherence for the squeezed 
relic photons does not go to $3$ in the large-scale limit but rather to $5/3$. 

It is interesting to stress, as we close, that the single-mode approximation is perfectly sound 
when the fluctuations beyond the horizon are described by a scalar field as it happens for the 
curvature perturbations \cite{mg2}. In this case we could even go to higher order and compute the 
degrees of third- or fourth-order coherence (see Eqs. (\ref{g3}) and (\ref{g4})) and confirm the same 
result. While the lengthy details will be omitted we can say that $\overline{g}^{(3)}=11 + {\mathcal O}(1/\overline{n})$ and $\overline{g}^{(4)} 
= 93 +  {\mathcal O}(1/\overline{n})$: this result holds also in the scalar case when all the modes of the field are taken into account.
In the case of the squeezed relic photons, however, the role of the polarizations 
is essential, as the comparison of Eqs. (\ref{B2result}), (\ref{S2result}) and (\ref{expl2A}) clearly shows. 

\renewcommand{\theequation}{6.\arabic{equation}}
\setcounter{equation}{0}
\section{Concluding remarks}
\label{sec6}
Among the six fiducial parameters characterizing the concordance scenario with massless neutrinos,  a single 
number (i.e. the scalar spectral index) accounts for the presence of large-scale inhomogeneities. 
A further source of inhomogeneity is represented by the tensor modes of the 
geometry even if their amplitude is, at least, one order of magnitude smaller than the one of the scalar modes.
Furthermore since we do observe magnetic fields over large distance scales we may even admit
the presence of large-scale gauge inhomogeneities. 
In the standard lore provided by conventional inflationary models all the potential sources 
of large-scale perturbations could stem from the zero-point fluctuations 
of quantum fields of different spins. At the moment the only argument in 
favour for this appealing possibility is merely theoretical: since a long stage of inflation is 
supposed to iron efficiently all preexisting inhomogeneities, it is logically plausible 
that large-scale fluctuations originated quantum mechanically. Because of the various 
assumptions behind this  suggestion, it would be highly desirable to a have 
a more operational way of deciding about the statistical properties of large-scale 
fluctuations. 

As we showed a possible answer to these questions involves the application 
of the tenets of Glauber theory, originally developed to address the coherence properties of
 optical fields. This 
analysis can be applied to the large-scale curvature perturbations but also 
to the large-scale fluctuations of the gauge fields. Since the pioneering attempts of Hanbury Brown and Twiss, it 
has been realized that the study of first order interference 
between the amplitudes cannot be used to distinguish the nature of different quantum states of the radiation field.
Young interferometry (indirectly based on the concept of power spectrum) is not able, by itself, 
to provide information on the statistical properties of the quantum correlations 
since various states with diverse physical properties (such as laser light and chaotic light) may lead 
to comparable degrees of first-order coherence. It is only by correlating intensities that the possible 
quantum origin of large-scale inhomogeneities can be independently assessed.
 In quantum optics the Glauber approach is often used in an exclusive manner by reducing the statistical 
properties of light to the analysis of a single (polarized) mode of the field: this is commonly referred to as
 the single-mode approximation. When dealing with large-scale fluctuations of different spins 
in cosmology the approach can only be inclusive since the correlation functions are typically unpolarized 
and contain all the modes of the field.

While the overall attempt of this paper is rather pragmatic, the obtained results are 
potentially inspiring. The modest viewpoint conveyed in this analysis is that precision cosmology, 
by itself, cannot validate its own premises. If new generations of astrophysical detectors will be able to resolve 
single photons the analysis of second-order interference effects may become feasible, at least in the case 
of the Cosmic Microwave Background.

\section*{Acknowledgements}
It is a pleasure to thank T. Basaglia and S. Rohr of the CERN scientific information service for their kind assistance.  

\newpage
\begin{appendix}
\renewcommand{\theequation}{A.\arabic{equation}}
\setcounter{equation}{0}
\section{Basic conventions and notations}
\label{APPA}
In time-dependent conformally flat backgrounds and in the Coulomb gauge (i.e. $Y_{0} =0$  and $\vec{\nabla}\cdot \vec{Y} =0$) 
the action (\ref{action}) can be written as:
\begin{equation}
S = \frac{1}{2}\int \,d\tau\, d^{3} x \, \biggl\{ \vec{{\mathcal A}}^{\,\prime \,2} + \biggl(\frac{\chi_{E}^{\,\prime}}{\chi_{E}}\biggr)^2 
 \vec{{\mathcal A}}^{\,2}  - 2 \frac{\chi_{E}'}{\chi_{E}} \vec{{\mathcal A}} \cdot \vec{{\mathcal A}}^{\,\prime} - \frac{\chi_{B}^2}{\chi_{E}^2}\partial_{i} \vec{{\mathcal A}} \cdot \partial^{i} \vec{{\mathcal A}}\biggr\},
\label{actold}
\end{equation}
where\footnote{The $1/\sqrt{4\pi}$ is purely conventional 
and its presence comes from the factor $16\pi$ included in the initial gauge action. } $\vec{{\mathcal A}} = \sqrt{ \Lambda_{E}/(4\pi)} \vec{Y}$; we have assumed that $\chi_{E}$ and $\chi_{B}$ are only dependent on the conformal time coordinate $\tau$. In terms of the canonical momentum $\vec{\pi}$ conjugate to $\vec{{\mathcal A}}$ 
the canonical Hamiltonian is simply given by:
\begin{equation}
H(\tau) = \frac{1}{2} \int d^3 x \biggl[ \vec{\pi}^{2} + 2 \frac{\chi_{E}'}{\chi_{E}} \vec{\pi} \cdot \vec{{\mathcal A}} + 
\frac{\chi_{B}^2}{\chi_{E}^2} \partial_{i} \vec{{\mathcal A}} \cdot \partial^{i} \vec{{\mathcal A}}\biggr], \qquad \vec{\pi} = \vec{{\mathcal A}}^{\,\prime} - \frac{\chi_{E}'}{\chi_{E}} \vec{{\mathcal A}}.
\label{hamold}
\end{equation}
The discussion can be carried on in the case of different susceptibilities and different gauge couplings (see e.g. \cite{action});
however we shall now focus on the case $\chi_{E} = \chi_{B} = \chi$ so that Eq. (\ref{hamold}) becomes:
\begin{equation}
H(\tau) = \frac{1}{2} \int d^3 x \biggl[ \vec{\pi}^{2} + 2 \frac{\chi'}{\chi} \vec{\pi} \cdot \vec{{\mathcal A}} + 
 \partial_{i} \vec{{\mathcal A}} \cdot \partial^{i} \vec{{\mathcal A}}\biggr], \qquad \vec{\pi} = \vec{{\mathcal A}}^{\,\prime} - \frac{\chi'}{\chi} \vec{{\mathcal A}}.
 \label{hamold2}
\end{equation}
The vector potential and the canonical momenta are explicitly related to the canonical 
electric and magnetic fields as  $\vec{B} = \vec{\nabla} \times \vec{{\mathcal A}}$ and  as $\vec{E} = - \vec{\pi}$.
In Fourier space the corresponding field operators are:
\begin{eqnarray}
\hat{{\mathcal A}}_{i}(\vec{x},\tau) &=& \frac{1}{\sqrt{V}} \sum_{\vec{p}, \, \alpha} e^{(\alpha)}_{i} \, \hat{{\mathcal A}}_{\vec{p},\, \alpha}(\tau)\, e^{- i \vec{p}\cdot \vec{x}},\qquad  \hat{{\mathcal A}}_{\vec{p},\, \alpha} = \frac{1}{\sqrt{2 p}} ( \hat{a}_{\vec{p}\, \alpha} + \hat{a}_{-\vec{p}\, \alpha}^{\dagger}),
\label{AA2}\\
\hat{\pi}_{i}(\vec{x},\tau) &=& \frac{1}{\sqrt{V}} \sum_{\vec{p}, \, \alpha} e^{(\alpha)}_{i} \,\hat{\pi}_{\vec{p}\, \alpha}(\tau)\, e^{- i \vec{p}\cdot \vec{x}},
\qquad \hat{\pi}_{\vec{p},\, \alpha} = - i \sqrt{\frac{p}{2}} ( \hat{a}_{\vec{p}\, \alpha} - \hat{a}_{-\vec{p}\, \alpha}^{\dagger}),
\label{AA3}
\end{eqnarray}
where $V$ is a fiducial (normalization) volume. In the discussion it is practical 
to switch from discrete to continuous modes where the creation 
and annihilation operators obey $[\hat{a}_{\vec{k}\, \alpha}, \hat{a}_{\vec{p}\, \beta}^{\dagger}] = \delta_{\alpha\beta}
\delta^{(3)}(\vec{k} - \vec{p})$ and the sums are replaced by integrals according to 
$\sum_{\vec{k}} \to V \int d^{3} k/(2\pi)^3$. This observation should be borne in mind when discussing 
the explicit results; in terms of Eqs. (\ref{AA2}) and (\ref{AA3}) the Hamiltonian of Eq. (\ref{hamold2}) becomes exactly the one reported in Eq. (\ref{one}).
\renewcommand{\theequation}{B.\arabic{equation}}
\setcounter{equation}{0}
\section{Four-point functions}
\label{APPB}
We report here some of the explicit expressions involved in the derivations of the four-point functions appearing in sections \ref{sec4} and \ref{sec5}. Let us recall that, according to the present 
conventions:
\begin{eqnarray}
\hat{E}_{i}^{(-)}(\vec{x}, \tau) &=& - \frac{i}{\sqrt{V}} \sum_{\vec{p},\,\alpha} \sqrt{\frac{p}{2}}  e^{(\alpha)}_{i} \hat{a}_{-\vec{p}, \, \alpha}^{\dagger} e^{- i \vec{p}\cdot\vec{x}}, \,\,\,\,\,
\hat{B}_{i}^{(-)}(\vec{x}, \tau)= - \frac{i}{\sqrt{V}} \, \sum_{\vec{p},\, \alpha} \frac{\epsilon_{m n i} \,  p _{m} \, e^{(\alpha)}_{n} }{\sqrt{2 p}} \hat{a}_{-\vec{p},\, \alpha}^{\dagger}\, e^{- i \vec{p} \cdot \vec{x}}
\nonumber\\
\hat{E}_{i}^{(+)}(\vec{x}, \tau) &=& \frac{i}{\sqrt{V}} \sum_{\vec{p},\,\alpha} \sqrt{\frac{p}{2}}  e^{(\alpha)}_{i} \hat{a}_{\vec{p}, \, \alpha} e^{- i \vec{p}\cdot\vec{x}},  \,\,\,\,\, \hat{B}_{i}^{(+)}(\vec{x}, \tau) = - \frac{i}{\sqrt{V}} \, \sum_{\vec{p},\, \alpha} \frac{\epsilon_{m n i} \,  p _{m} \, e^{(\alpha)}_{n} }{\sqrt{2 p}} \hat{a}_{\vec{p},\, \alpha} e^{- i \vec{p} \cdot \vec{x}}.
\nonumber
\end{eqnarray}
The two-point functions define the degree of first-order coherence and they are:
\begin{eqnarray}
{\mathcal E}^{(1)}(x_{1}, x_{2}) &=& \langle \hat{E}_{i}^{(-)}(x_{1}) \, \hat{E}^{(+)}_{i}(x_{2}) \rangle 
\nonumber\\
&=& \frac{1}{2 V} \sum_{\vec{p}_{1},\,\alpha_{1}} e^{- i \vec{p}_{1}\cdot \vec{x}_{1}} \sum_{\vec{p}_{2},\,\alpha_{2}} e^{- i \vec{p}_{2}\cdot \vec{x}_{2}} \sqrt{p_{1} p_{2}} 
\nonumber\\
&\times& e^{(\alpha_{1})}_{i}(\hat{p}_{1}) \, e^{(\alpha_{2})}_{i}(\hat{p}_{2})\,\, \langle \hat{a}^{\dagger}_{-\vec{p}_{1}, \alpha_{1}} \,  \hat{a}_{\vec{p}_{2}, \alpha_{2}} \rangle,
\nonumber\\
{\mathcal B}^{(1)}(x_{1}, x_{2}) &=& \langle \hat{B}_{i}^{(-)}(x_{1}) \, \hat{B}^{(+)}_{i}(x_{2}) \rangle
\nonumber\\
&=& - \frac{1}{2 V}  \sum_{\vec{p}_{1},\,\alpha_{1}} e^{- i \vec{p}_{1}\cdot \vec{x}_{1}} \sum_{\vec{p}_{2},\,\alpha_{2}} e^{- i \vec{p}_{2}\cdot \vec{x}_{2}} \frac{1}{\sqrt{p_{1} p_{2}} }\nonumber\\
&\times& \epsilon_{m_{1} n_{1} i} p_{1 m_{1}} e_{n_{1}}^{(\alpha_{1})}(\hat{p}_{1})\,\epsilon_{m_{2} n_{2} i} p_{2 m_{2}} e_{n_{2}}^{(\alpha_{2})}(\hat{p}_{2})  \langle \hat{a}^{\dagger}_{-\vec{p}_{1}, \alpha_{1}} \,  \hat{a}_{\vec{p}_{2}, \alpha_{2}} \rangle.
\label{B2}
\end{eqnarray}
Using Eq. (\ref{corr5}) in the case $n=2$ we have: 
\begin{eqnarray}
{\mathcal G}^{(2)}(x_{1},\,x_{2},\,x_{3},\, x_{4}) &=& \langle {\mathcal A}_{i}^{(-)}(x_{1})\,{\mathcal A}_{j}^{(-)}(x_{2}) {\mathcal A}_{i}^{(+)}(x_{3}) {\mathcal A}_{j}^{(+)}(x_{4}) \rangle
\nonumber\\
&=& \frac{1}{4 V^2} \sum_{\vec{p}_{1}, \,\alpha_{1}}\, \frac{e^{- i \vec{p}_{1}\cdot \vec{x}_{1}}}{\sqrt{p_{1}}}\,\sum_{\vec{p}_{2}, \,\alpha_{2}} \, 
\frac{e^{- i \vec{p}_{2}\cdot \vec{x}_{2}}}{\sqrt{p_{2}}}\, \sum_{\vec{p}_{3}, \,\alpha_{3}} \, \frac{e^{- i \vec{p}_{3}\cdot \vec{x}_{3}}}{\sqrt{p_{3}}}\, \sum_{\vec{p}_{4}, 
\,\alpha_{4}}  \frac{e^{- i \vec{p}_{4}\cdot \vec{x}_{4}}}{\sqrt{p_{4}}}\,
\nonumber\\
&\times& \, e^{(\alpha_{1})}_{i}(\hat{p}_{1}) \, e^{(\alpha_{2})}_{j}(\hat{p}_{2}) \,  e^{(\alpha_{3})}_{i}(\hat{p}_{3})\, e_{j}^{(\alpha_{4})}(\hat{p}_{4})
\nonumber\\
&\times& \langle \hat{a}^{\dagger}_{-\vec{p}_{1}, \alpha_{1}} \, \hat{a}^{\dagger}_{-\vec{p}_{2}, \alpha_{2}} \, \hat{a}_{\vec{p}_{3}, \alpha_{3}} \, \hat{a}_{\vec{p}_{4}, \alpha_{4}} \rangle.
\label{vector4}
\end{eqnarray}
The degrees of quantum coherence can also be defined in terms of the electric fields themselves, as 
originally suggested by Glauber.  Equation (\ref{corrE}) in the case 
$n=2$ becomes:
\begin{eqnarray}
{\mathcal E}^{(2)}(x_{1},\,x_{2},\,x_{3},\, x_{4}) &=& \langle E_{i}^{(-)}(x_{1})\,E_{j}^{(-)}(x_{2}) E_{i}^{(+)}(x_{3}) E_{j}^{(+)}(x_{4}) \rangle
\nonumber\\
&=& \frac{1}{4 V^2} \sum_{\vec{p}_{1}, \,\alpha_{1}}\, e^{- i \vec{p}_{1}\cdot \vec{x}_{1}}\,\sum_{\vec{p}_{2}, \,\alpha_{2}} \, e^{- i \vec{p}_{2}\cdot \vec{x}_{2}}\, \sum_{\vec{p}_{3}, \,\alpha_{3}} \, e^{- i \vec{p}_{3}\cdot \vec{x}_{3}}\, \sum_{\vec{p}_{4}, \,\alpha_{4}}  e^{- i \vec{p}_{4}\cdot \vec{x}_{4}}\,
\nonumber\\
&\times& \sqrt{p_{1}\, p_{2}\, p_{3}\, p_{4}}\, e^{(\alpha_{1})}_{i}(\hat{p}_{1}) \, e^{(\alpha_{2})}_{j}(\hat{p}_{2}) \,  e^{(\alpha_{3})}_{i}(\hat{p}_{3})\, e_{j}^{(\alpha_{4})}(\hat{p}_{4})
\nonumber\\
&\times& \langle \hat{a}^{\dagger}_{-\vec{p}_{1}, \alpha_{1}} \, \hat{a}^{\dagger}_{-\vec{p}_{2}, \alpha_{2}} \, \hat{a}_{\vec{p}_{3}, \alpha_{3}} \, \hat{a}_{\vec{p}_{4}, \alpha_{4}} \rangle.
\label{electric4}
\end{eqnarray}
Finally, if we write Eq. (\ref{corrB}) in the case $n=2$ 
the result is:
\begin{eqnarray}
{\mathcal B}^{(2)}(x_{1},\,x_{2},\,x_{3},\, x_{4}) &=& \langle B_{i}^{(-)}(x_{1})\,B_{j}^{(-)}(x_{2}) B_{i}^{(+)}(x_{3}) B_{j}^{(+)}(x_{4}) \rangle
\nonumber\\
&=& \frac{1}{4 V^2} \sum_{\vec{p}_{1}, \,\alpha_{1}}\, \frac{e^{- i \vec{p}_{1}\cdot \vec{x}_{1}}}{\sqrt{p_{1}}}\,\sum_{\vec{p}_{2}, \,\alpha_{2}} \, \frac{e^{- i \vec{p}_{2}\cdot \vec{x}_{2}}}{\sqrt{p_{2}}}\, \sum_{\vec{p}_{3}, \,\alpha_{3}} \, \frac{e^{- i \vec{p}_{3}\cdot \vec{x}_{3}}}{\sqrt{p_{3}}}\, \sum_{\vec{p}_{4}, \,\alpha_{4}}  \frac{e^{- i \vec{p}_{4}\cdot \vec{x}_{4}}}{\sqrt{p_{4}}}\,
\nonumber\\
&\times& \, e^{(\alpha_{1})}_{n_{1}}(\hat{p}_{1}) \epsilon_{m_{1} n_{1} i} \,p_{1, m_{1}} \, e^{(\alpha_{2})}_{n_{2}}(\hat{p}_{2}) \, \epsilon_{m_{2} n_{2} j} \,p_{2, m_{2}}  
\nonumber\\
&\times& e^{(\alpha_{3})}_{n_{3}}(\hat{p}_{3})\,\epsilon_{m_{3} n_{3} i} \,p_{3, m_{3}}  \, e_{n_{4}}^{(\alpha_{4})}(\hat{p}_{4}) \epsilon_{m_{4} n_{4} j} \,p_{4, m_{4}} 
\nonumber\\
&\times& \langle \hat{a}^{\dagger}_{-\vec{p}_{1}, \alpha_{1}} \, \hat{a}^{\dagger}_{-\vec{p}_{2}, \alpha_{2}} \, \hat{a}_{\vec{p}_{3}, \alpha_{3}} \, \hat{a}_{\vec{p}_{4}, \alpha_{4}} \rangle.
\label{magnetic4}
\end{eqnarray}
To compute the degree of second-order coherence we need the following expectation value: 
\begin{eqnarray}
&&\langle \hat{a}^{\dagger}_{-\vec{p}_{1}, \alpha_{1}} \, \hat{a}^{\dagger}_{-\vec{p}_{2}, \alpha_{2}} \, \hat{a}_{\vec{p}_{3}, \alpha_{3}} \, \hat{a}_{\vec{p}_{4}, \alpha_{4}} \rangle =
\nonumber\\
&&v_{p_{1}}^{*}(\tau_{1}) v_{p_{2}}^{*}(\tau_{2}) v_{p_{3}}(\tau_{3})  v_{p_{4}}(\tau_{4}) \biggl[ \delta^{(3)}(\vec{p}_{1} + \vec{p}_{4})  \delta^{(3)}(\vec{p}_{2} + \vec{p}_{3})\delta_{\alpha_{1}\alpha_{4}} \delta_{\alpha_{2} \alpha_{3}} 
\nonumber\\
&&+ \delta^{(3)}(\vec{p}_{1} + \vec{p}_{3})  \delta^{(3)}(\vec{p}_{2} + \vec{p}_{4})
\delta_{\alpha_{1}\alpha_{3}} \delta_{\alpha_{2} \alpha_{4}}\biggr]
\nonumber\\
&&+ v_{p_{1}}^{*}(\tau_{1}) u_{p_{2}}^{*}(\tau_{2}) u_{p_{3}}(\tau_{3})  v_{p_{4}}(\tau_{4}) \delta^{(3)}(\vec{p}_{1} + \vec{p}_{2})  \delta^{(3)}(\vec{p}_{3} + \vec{p}_{4})\delta_{\alpha_{1}\alpha_{2}} \delta_{\alpha_{3} \alpha_{4}}.
\label{magnetic5}
\end{eqnarray}
It is important to contrast the results obtained in the vector case with the scalar case. 
\begin{eqnarray}
{\mathcal S}^{(2)}(x_{1},\,x_{2},\,x_{3},\, x_{4}) &=& \langle q^{(-)}(x_{1})\,q^{(-)}(x_{2}) q^{(+)}(x_{3}) q^{(+)}(x_{4}) \rangle
\nonumber\\
&=& \frac{1}{4 V^2} \sum_{\vec{p}_{1}, \,\alpha_{1}}\, e^{- i \vec{p}_{1}\cdot \vec{x}_{1}}\,\sum_{\vec{p}_{2}, \,\alpha_{2}} \, e^{- i \vec{p}_{2}\cdot \vec{x}_{2}}\, \sum_{\vec{p}_{3}, \,\alpha_{3}} \, e^{- i \vec{p}_{3}\cdot \vec{x}_{3}}\, \sum_{\vec{p}_{4}, \,\alpha_{4}}  e^{- i \vec{p}_{4}\cdot \vec{x}_{4}}\,
\nonumber\\
&\times& \sqrt{p_{1}\, p_{2}\, p_{3}\, p_{4}}\, 
\nonumber\\
&\times& \langle \hat{d}^{\dagger}_{-\vec{p}_{1}} \, \hat{d}^{\dagger}_{-\vec{p}_{2}} \, \hat{d}_{\vec{p}_{3}} \, \hat{d}_{\vec{p}_{4}} \rangle.
\label{scalar1}
\end{eqnarray}
where, in this case, 
\begin{eqnarray}
\langle \hat{d}^{\dagger}_{-\vec{p}_{1}} \, \hat{d}^{\dagger}_{-\vec{p}_{2}} \, \hat{d}_{\vec{p}_{3}} \, \hat{d}_{\vec{p}_{4}} \rangle &=&
v_{p_{1}}^{*}(\tau_{1}) v_{p_{2}}^{*}(\tau_{2}) v_{p_{3}}(\tau_{3})  v_{p_{4}}(\tau_{4}) \biggl[ \delta^{(3)}(\vec{p}_{1} + \vec{p}_{4})  \delta^{(3)}(\vec{p}_{2} + \vec{p}_{3})
\nonumber\\
&+& \delta^{(3)}(\vec{p}_{1} + \vec{p}_{3})  \delta^{(3)}(\vec{p}_{2} + \vec{p}_{4})\biggr]
\nonumber\\
&+& v_{p_{1}}^{*}(\tau_{1}) u_{p_{2}}^{*}(\tau_{2}) u_{p_{3}}(\tau_{3})  v_{p_{4}}(\tau_{4}) 
\delta^{(3)}(\vec{p}_{1} + \vec{p}_{2})  \delta^{(3)}(\vec{p}_{3} + \vec{p}_{4}).
\label{scalar2}
\end{eqnarray}
As already mentioned after Eq. (\ref{AA3}) in the continuous mode representation we have that 
the commutation relations are $[\hat{a}_{\vec{k}\, \alpha}, \hat{a}_{\vec{p}\, \beta}^{\dagger}] = \delta_{\alpha\beta}
\delta^{(3)}(\vec{k} - \vec{p})$. Clearly in the discrete mode representation the commutation relations 
will contain the appropriate volume factors and the Dirac delta functions will be replaced by Kroeneker 
deltas over the discrete momenta. The two procedures are fully equivalent.
\renewcommand{\theequation}{C.\arabic{equation}}
\setcounter{equation}{0}
\section{Power spectra}
\label{APPC}
The power spectra when the relevant scales are larger than the Hubble radius and before reentry are given by:
\begin{eqnarray}
P_{B}(k,\tau) &=& \frac{\pi}{2} C^2 |H_{\nu}^{(1)}(x_{i})|^2 
\nonumber\\
&+& \frac{\pi}{2 \beta} \biggl[ H^{(1)}_{\nu}(x_{i})\,H^{(2)}_{\nu-1}(x_{i}) + H^{(1)}_{\nu-1}(x_{i}) H^{(2)}_{\nu}(x_{i})\biggr]\biggl(D + 2 \ln{x_{i}} - \ln{D \beta^2}\biggr) x_{i}^2
+ {\mathcal O}(x_{i}^3), 
\nonumber\\
P_{E}(k,\tau) &=& \frac{\pi}{2 C^2} |H^{(1)}_{\nu-1}(x_{i})|^2\, x_{i} + {\mathcal O}(x_{i}).
\label{corr}
\end{eqnarray}
If the evolution of the susceptibility is not continuous (or not differentiable) we can still write 
a generic form of the $u_{k}(\tau)$ and $v_{k}(\tau)$, namely:
\begin{eqnarray}
u_{k}(\tau) - v_{k}^{*}(\tau) &=& c_{-}(x_{i}) e^{i k (\tau + \delta_{k} \tau_{i})} +  c_{+}(x_{i}) e^{-i k (\tau + \delta_{k} \tau_{i})}, 
\nonumber\\
u_{k}(\tau) + v_{k}^{*}(\tau) &=&   c_{+}(x_{i}) e^{-i k (\tau + \delta_{k} \tau_{i})} - c_{-}(x_{i}) e^{i k (\tau + \delta_{k} \tau_{i})},
\end{eqnarray}
where $\delta_{k}$, in this context, is just an arbitrary phase possibly picked up at the transition and, as usual, $x_{i} = k \tau_{i}$. 
While in principle $c_{\pm}(x_{i})$ cannot be determined since the evolution is not continuous 
we can try to fix them by imposing, artificially, the continuity of the solutions for $\tau< - \tau_{i}$ and $\tau \geq - \tau_{i}$.
The result of this procedure will be 
\begin{eqnarray}
\overline{c}_{-}(x_{i}) &=& \frac{{\mathcal N}}{2} \sqrt{ x_{i}} \biggl[ H_{\nu}^{(1)}(x_{i}) + i  H_{\nu-1}^{(1)}(x_{i}) \biggr] e^{- i \delta_{k} x_{i}},
\label{cc1}\\
\overline{c}_{+}(x_{i}) &=& \frac{{\mathcal N}}{2} \sqrt{ x_{i}} \biggl[ H_{\nu}^{(1)}(x_{i}) - i  H_{\nu-1}^{(1)}(x_{i}) \biggr] e^{ i \delta_{k} x_{i}}.
\label{cc2}
\end{eqnarray}
where, for simplicity, we denoted $c_{\mp}(x_{i}) = \overline{c}_{\mp}(x_{i}) e^{\mp i \delta_{k}}$. 
The magnetic and the electric power spectra are, respectively, 
\begin{equation}
P_{B}(k,\tau) =\frac{k^4}{4 \pi^2} |u_{k}(\tau) - v_{k}^{*}(\tau)|^2, \qquad P_{E}(k,\tau) =\frac{k^4}{4 \pi^2} |u_{k}(\tau) + v_{k}^{*}(\tau)|^2 
\label{PEone}
\end{equation}
Using Eqs. (\ref{cc1}) and (\ref{cc2}), Eq. (\ref{PEone}) becomes:
\begin{eqnarray}
P_{B}(k,\, x_{i},\,\tau) &=& \frac{H_{i}^4  a_{i}^4 x_{i}^5}{8 \pi} \biggl\{ \biggl[ J_{\nu}(x_{i}) \cos{k \tau} - J_{\nu -1}(x_{i}) \sin{k\tau}\biggr]^2 
\nonumber\\
&+& \biggl[ Y_{\nu}(x_{i}) \cos{k \tau} - Y_{\nu -1}(x_{i}) \sin{k\tau}\biggr]^2\biggr\},
\nonumber\\
 P_{E}(k,\, x_{i},\, \tau) &=& \frac{H_{i}^4  a_{i}^4 x_{i}^5}{8 \pi} \biggl\{ \biggl[ J_{\nu}(x_{i}) \sin{k \tau} + J_{\nu -1}(x_{i}) \cos{k\tau}\biggr]^2 
\nonumber\\
&+& \biggl[ Y_{\nu}(x_{i}) \sin{k \tau} + Y_{\nu -1}(x_{i}) \cos{k\tau}\biggr]^2\biggr\}.
\label{PBoneb}
\end{eqnarray}
In the sudden approximation (i.e. $\beta \to \infty$ and $C\to 1$) Eqs. (\ref{corr}) and (\ref{PBoneb}) give the 
same result for $x_{i} \ll1 $ and $k\tau < 1$. The reverse is not always true since the technique leading to Eq. (\ref{PBoneb})
is based on the continuity of the susceptibility which is not verified in practice. The correct junction conditions 
for the susceptibility and for the extrinsic curvature are therefore essential for a correct derivation 
of the power spectra and of the degrees of quantum coherence.
\end{appendix}

\end{document}